\documentclass{emulateapj}

\newcommand{\ee}[1]{\mbox{${} \times 10^{#1}$}}
\newcommand{\eten}[1]{\mbox{$10^{#1}$}}

\newcommand{\h}{\mbox{$^h$}}
\newcommand{\m}{\mbox{$^m$}}
\newcommand{\s}{\mbox{$^s$}}
\newcommand{\degree}{\mbox{$^{\circ}$}}
\newcommand{\am}{\mbox{\arcmin}}
\newcommand{\as}{\mbox{\arcsec}}

\newcommand{\kms}{\mbox{km s$^{-1}$}}
\newcommand\cmv{\mbox{cm$^{-3}$}}
\newcommand\cmc{\mbox{cm$^{-2}$}}

\newcommand\submm{submillimeter}

\newcommand{\tk}{\mbox{$T_K$}}
\newcommand{\td}{\mbox{$T_d$}}

\newcommand{\mean}[1]{\mbox{$\langle#1\rangle$}} 

\newcommand{\form}{H$_2$CO}

\newcommand{\ammonia}{\mbox{{\rm NH}$_3$}}
\newcommand{\nthp}{\mbox{{\rm N$_2$H}$^+$}}

\input{epsf}

\begin{document}

\title {\bf Probing Pre-Protostellar Cores with
Formaldehyde}
\author{Kaisa E. Young, Jeong-Eun Lee, Neal J. Evans II}
\affil{Department of Astronomy, The University of Texas at Austin,
       1 University Station C1400, Austin, Texas 78712--0259}
\author{Paul F. Goldsmith} 
\affil{Department of Astronomy, Cornell University, 522 Space Sciences 
       Building, Ithaca, New York 14853}
\author{Steven D. Doty} 
\affil{Department of Physics and Astronomy, Denison University, 
	Granville, Ohio 43023}


\begin{abstract}

We present maps of the 6 cm and 1.3 mm transitions of H$_2$CO toward
three cold, dense pre-protostellar cores: L1498, L1512, and L1544. The
6 cm transition is a unique probe of high density gas at low
temperature. However, our models unequivocally indicate that \form\ is
depleted in the interiors of PPCs, and depletion significantly affects
how \form\ probes the earliest stages of star formation. Multi-stage,
self-consistent models, including gas--dust energetics, of both \form\
transitions are presented, and the implications of the results are
discussed.

\end{abstract}



\section{Introduction}

Molecular cloud cores that have yet to form a star provide the best
opportunity to determine the initial conditions for star formation.
However, the earliest evolution of a cloud core is very poorly
constrained, and its evolutionary track is unclear.  Recently, a
number of compact objects without IRAS sources but with considerable
\submm\ continuum emission have been identified (Ward-Thompson et al.\
1994).  Since these have no obvious source of internal luminosity but
are characterized by high column density, they are plausible
candidates for the stages just before formation of a central
object. These objects are classified as Pre-Protostellar Cores, or
PPCs.

Several of these objects have been mapped in \submm\ continuum
emission (Shirley et al.\ 2000), and the dust emission has been
modeled using self-consistent calculations of the dust temperature
(Evans et al.\ 2001).  The calculations show that the dust deep inside
the PPCs is very cold ($\td \sim 8$ K) and that this low \td\
affects the interpretation of the \submm\ emission. In particular, the
PPCs may be more centrally condensed, with higher central densities,
than previously thought.  Study of these objects will be very valuable
for our understanding of the formation of solar-type stars.  The
density distribution as a function of radius in such starless cores is
a strong discriminator between theoretical models. The density
structure is also of critical importance for assessing whether the
cloud is in equilibrium, or whether it may be beginning to collapse,
and the density structure of PPCs strongly affects the nature of the
collapse and star formation that follows (Andr\'{e} et al.\ 2000).

Most density tracers employed by astronomers involve the use of a pair
of rotational transitions of a linear molecule (e.g., isotopologues of
CO, CS, HC$_3$N).  Since the different transitions have different
spontaneous decay rates, they require different collision rates to
bring their populations into thermal equilibrium. At a specific
density, their relative populations reflect the collision rate and
thus allow determination of the hydrogen density. This method can work
well, but there is an important caveat: the different rotational
levels generally lie at significantly different energies above the
ground state, given approximately by the rigid--rotor formula $E(J) =
hB_0J(J + 1)$.  For C$^{18}$O, $J = 1$ is at 5.3 K, $J = 2$ is at 15.8
K, $J = 3$ at 31.6 K, etc.  When we are dealing with a very cold cloud
(T $\leq$ 10 K) or a cloud whose temperature structure is unknown, a
significant uncertainty is introduced.  The collisional excitation
rates, as well as the LTE level populations, for a linear molecule
having a large rotation constant, including C$^{18}$O, depend
sensitively on the kinetic temperature of the gas.  The lack of
detailed temperature information means that the density one determines
becomes confused with the temperature.  It is extremely difficult to
disentangle the two, especially in the case of cold dark clouds which
may have warm edges heated by the interstellar radiation field.  For
PPCs, the heated edge can dominate the excitation effects and only
with quite detailed models can we become confident that we have
properly separated the temperature and the density structure.

\begin{figure}
\plotone{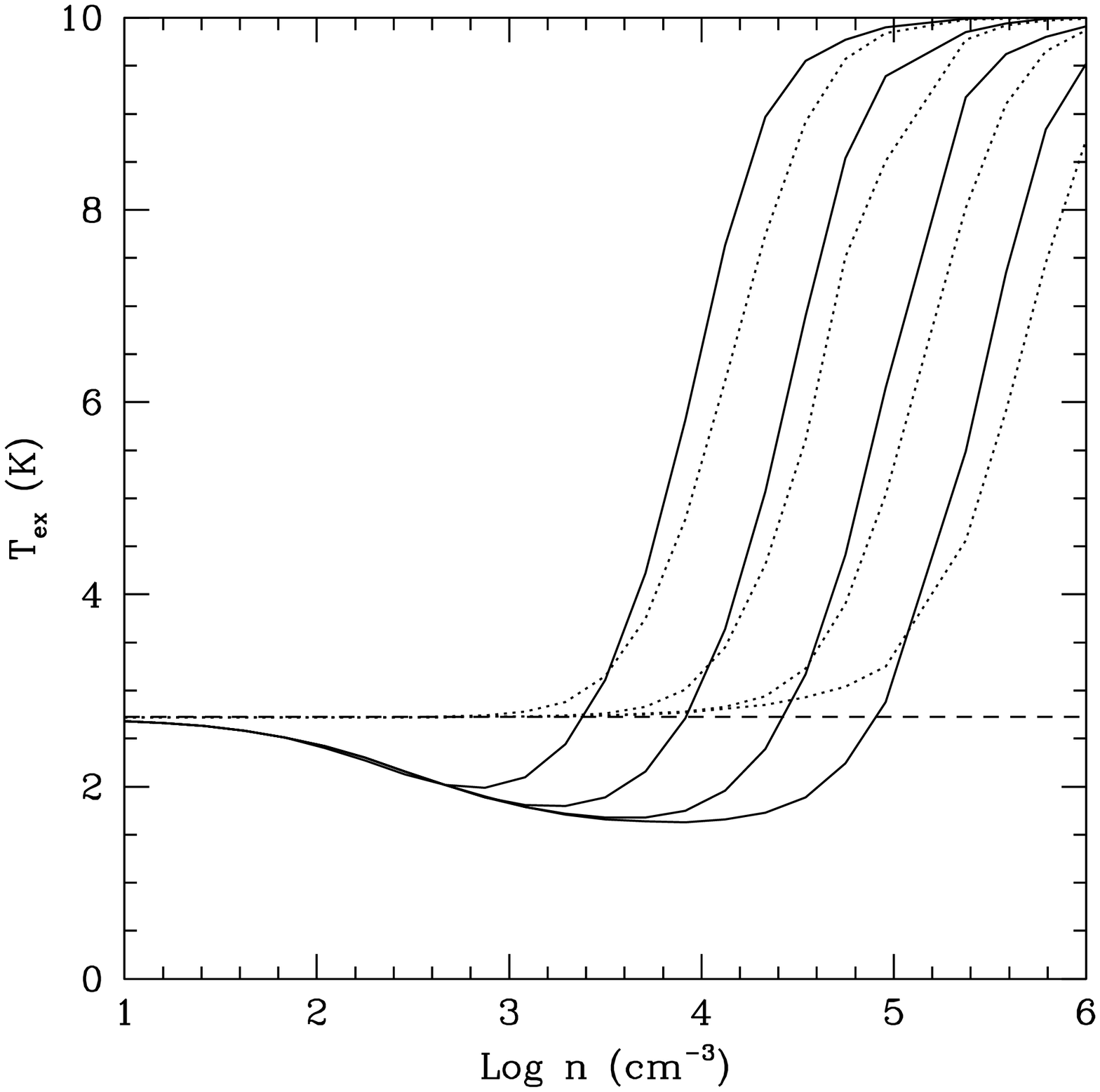} \figcaption{\label{tex} Plot of the excitation
temperature of the H$_2$CO 6 cm (solid lines) and 1.3 mm (dotted
lines) transitions at different abundances versus the log of the
molecular hydrogen density. The \form\ fractional abundance decreases
from left to right for the curves in the plot, $X = 10^{-7}, 10^{-8},
10^{-9}, 10^{-10}$.  The dashed line is the temperature of the CBR
(2.725 K). The plot was generated assuming $T_K$ = 10 K, using an LVG
code with $dv/dr = 1$ km s$^{-1}$ pc$^{-1}$.}
\end{figure}

Clearly a better probe of density for studying cold, dark clouds would
be valuable.  The lowest transition of ortho--formaldehyde (H$_2$CO)
at 6 cm is one possible candidate. The 6 cm line is seen in absorption
against the cosmic background radiation (CBR) in molecular clouds.
The discovery of absorption against the cosmic background radiation by
the lowest transition of \form\ (Zuckerman et al.\ 1969) was quite
surprising because it is impossible for any two--level system to have
an excitation temperature below that of the CBR. The 6 cm transition
occurs between the levels of the lowest K-doublet of \form\ (the
$J_{K_{-1}K_{1}}$ = $1_{11}$ and $1_{10}$ levels). Townes and Cheung
(1969) proposed a collisional pumping scheme to lower the excitation
temperature; this scheme depends upon collisions to higher rotational
levels followed by radiative decay to the $J = 1$ levels.  The
propensity of collisions to favor the lower levels in each excited
K-doublet transition leads to an overpopulation of the lower level of
the $J=1$ K-doublet. This pumping scheme operates over a range of
densities up to about $\eten4$ \cmv, at which point collisions become
dominant and drive the excitation temperature toward the kinetic
temperature. Thus, there is a range of densities where $T_{ex} \sim
T_{CBR}$ with no emission or absorption (Figure \ref{tex}). At
even higher densities, the line goes into emission. Because of
trapping, the density above which the 6 cm line goes into emission
depends on abundance. The 6 cm line could act as a unique tracer of
cold, dense gas, because it does not require high temperatures to
achieve significant fractional population of the upper level, but goes
into emission {\it solely} as a result of high density.

The PPCs in this study are known to have high molecular hydrogen
central densities ($n_c = 10^4 - 10^6$ \cmv; Evans et al.\ 2001) and,
since there is no internal heat source, they should be very
cold. Because the 6 cm transition goes from absorption to emission in
these conditions (Figure\ \ref{tex}), one might expect emission, weak
absorption, or possibly neither toward the center of the PPCs. Vanden
Bout, Snell, and Wilson (1983) observed deep 6 cm \form\ absorption
toward seven dark clouds. That study concluded that the 6 cm line was
only probing the less dense outer envelope of the clouds.

Previous observations of the 6 cm \form\ transition, including those
of Vanden Bout et al.\ (1983), were carried out mostly with
very large beams ($\sim 3\arcmin$ to $6\arcmin$), with the exception
of a few observations with the VLA. In a VLA study of B335 (Zhou et
al.\ 1990), a dark cloud in the early phases of forming a low mass
star, the authors found a ring of absorption surrounding the center of
the object. They proposed that the absence of absorption toward the
center was caused by elevated densities near the center of the source.
This proposal was verified by observations of the mm lines with the
IRAM 30-m telescope, which in turn led to the identification of B335
as a collapsing protostar (Zhou et al.\ 1993). Additionally, the 6 cm
line of \form\ has been observed in emission toward regions of higher
mass star formation, such as Orion (Zuckerman, Rickard, \& Palmer
1975).

In this paper, we investigate this issue through high angular
resolution observations of the 6 cm line in three PPCs.  Sections
\ref{sec_obs} -- \ref{sec_res} describe the observations and the
results. Observations of an additional \form\ transition at 1.3 mm are
also described. In Section \ref{sec_mod}, we describe the methods used
to model the line profiles and show how the observations can be
explained by inclusion of depletion. These models are constrained by
the additional modeling of the 1.3 mm \form\ emission line observed
toward the PPCs. Section \ref{sec_dis} discusses the model results. A
summary is provided in Section \ref{sec_sum}. A detailed discussion of
the gas--dust energetics calculations is given in the Appendix.

\section{Observations}\label{sec_obs} 

Three PPCs, L1498 (04\h 10\m 51.5\s\ 25\degree 09\am 58\as, J2000),
L1512 (05\h 04\m 08.2\s\ 32\degree 43\am 20\as), and L1544 (05\h 04\m
17.1\s\ 25\degree 10\am 48\as), were observed in December 2001 and
September 2002 using the 305-m telescope of the Arecibo Observatory.
All of the sources were observed at 6 cm, the $J_{K_{-1}K_{1}} =
1_{11} - 1_{10}$ transition of ortho-formaldehyde.  The observed
frequency was 4829.6594 MHz, the weighted mean of the frequencies of
the main hyperfine component ($F=2-2$) and the nearby $F=0-1$
component as given in Tucker, Tomasevich, \& Thaddeus (1971). The
Arecibo telescope has a beam size of 60\arcsec\ at this frequency.  We
observed at least 15 positions for each of three PPCs in the 6 cm
line, creating maps with spectra spaced at 60\arcsec\ intervals.  The
mapping strategy ensured mapping off the edge of the core as seen in
the dust. The long direction of each map is perpendicular to the long
axis of the dust core. For L1544 and L1498, the long axis runs NW--SE;
therefore, the map runs NE--SW.

The first challenge was to develop an appropriate data taking
strategy.  The standard Arecibo position switching routine, with off
position designed to have the telescope observe the same ``track''
across the sky, is problematic, as these sources are large, and the
reference position could ``contaminate'' the data.  Switching to a
fixed reference position was an option but would have had negative
implications for data reduction.  Fortunately, taking total power ``ON
SOURCE'' spectra revealed that the receiver and autocorrelator (with
digital bandpass filters) gave a baseline that was limited only by the
fluctuations expected from the radiometer equation.  This method also
gains a factor of two in sensitivity compared to position switching,
and should be of general value for observation of narrow spectral
lines.  As discussed in Section 3, this technique worked well for
integration times as long as 200 min, resulting in rms fluctuations
below 10 mK in a 1.5 kHz channel width.

Standard Arecibo calibration sources were observed at the beginning
and end of each shift.  In December 2001, the average beam efficiency,
$\langle\eta_{B}\rangle$, including the main beam and the first
sidelobe, from these observations was 0.61$\pm$0.05. In September
2002, $\langle\eta_{B}\rangle=0.59\pm0.04$. This beam efficiency,
$\eta_{B}$, accounts for radiation (ohmic) losses, rearward and
forward scattering, and spillover. Assuming that the coupling
efficiency is unity because of the extended nature of the dark clouds,
we calculate the radiation (source) temperature, $T_R$, using the
equation
\begin{equation}
T_R = T_A/\eta_B,
\end{equation}
where $T_A$ is the measured antenna temperature.  $T_R$
is outside the atmosphere.  Table \ref{ppctab} lists the observed and
derived 6 cm line properties, including $T_R$.

In addition to the 6 cm observations, each PPC was observed with the
10.4-m Caltech Submillimeter Observatory (CSO)\footnote{The CSO is
operated by the California Institute of Technology under funding from
the National Science Foundation, contract AST 90-15755.} at 225.697787
GHz (1.3 mm), the frequency of the $J_{K_{-1}K_1} = 3_{12} - 2_{11}$
transition of \form. The CSO has a beamsize of 32\arcsec\ at 1.3
mm. The central position of L1544 was observed in October 1996. In
January 2002 and September/October 2003, the central positions of
L1498 and L1512 were observed, and five-point maps of all three PPCs
were made. The four offset positions were along and perpendicular to
the long axis of the core, all at 60\arcsec\ from the central position.

Observations of Mars and Jupiter as calibration sources gave an
average main beam efficiency at 230 GHz of
$\langle\eta_{MB}\rangle$=0.57$\pm$0.11 in October 1996,
$\langle\eta_{MB}\rangle$=0.53$\pm$0.04 in January 2002, and
$\langle\eta_{MB}\rangle$=0.77$\pm$0.06 in September/October
2003. Because the CSO employs a chopper wheel for calibration, the
measured quantity is $T_A^*$, the antenna temperature corrected for
atmospheric attenuation, radiation losses, and rearward spillover and
scatter. Therefore, $\eta_{MB}$ is $\eta_{B}$ divided by $\eta_{rss}$,
the ohmic and backward scattering efficiency.  The radiation
temperature in this case, again assuming the coupling efficiency is
unity, is given by
\begin{equation}
T_R = T_A^*/\eta_{MB}.
\end{equation}

All data were reduced using the Continuum Line Analysis Single-dish
Software, CLASS. CLASS was used to remove baselines, average weighted
spectra, and calibrate the data. The spectra were weighted by
$\sigma^{-2}$, where $\sigma$ is the rms noise. Spectra from different
nights were averaged. The 6 cm data were then Hanning smoothed. The
line center velocities ($v_{LSR}$), linewidths ($\Delta v$), and
optical depths ($\tau$) for the 6 cm data were determined by fitting a
manifold of Gaussian profiles, one for each hyperfine component. Only
one Gaussian component was necessary for the 1.3 mm data, because the
hyperfine structure is negligible due to the extremely close spacing
of the components.

\section{Results}\label{sec_res}  

\begin{figure}
\plotone{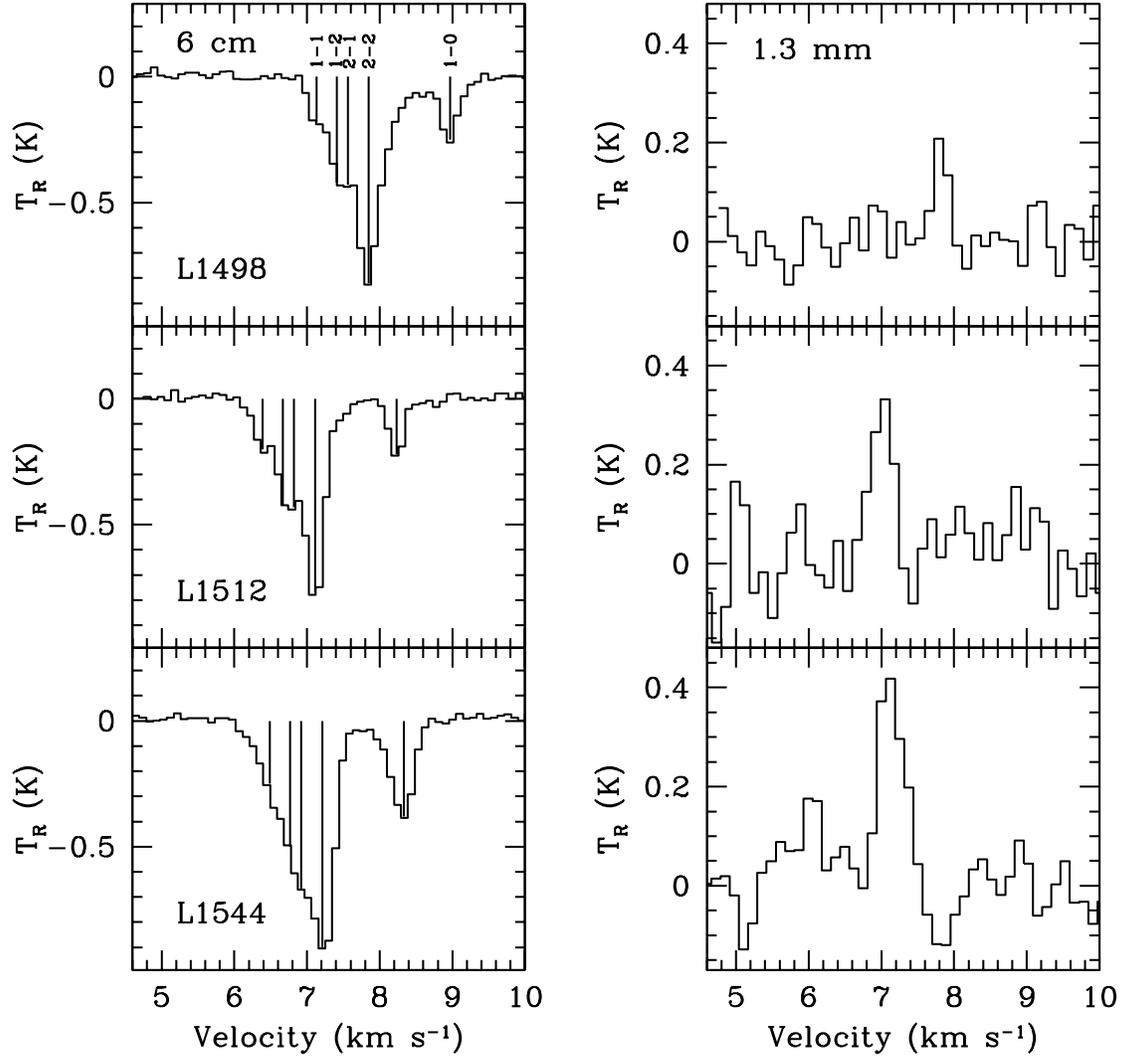} \figcaption{\label{centers} Spectra of
H$_2$CO in the dark clouds L1498, L1512, and L1544. Panels in the left
column show the 6 cm absorption against the CBR. The 6 cm hyperfine
components are marked with vertical lines and labeled in the top
panel. The $F = 0-1$ component is not marked, because it is blended
with the strongest component, $F = 2-2.$ Panels in the right column
show the 1.3 mm emission toward the centers of the PPCs.}
\end{figure}

\begin{figure}
\plotone{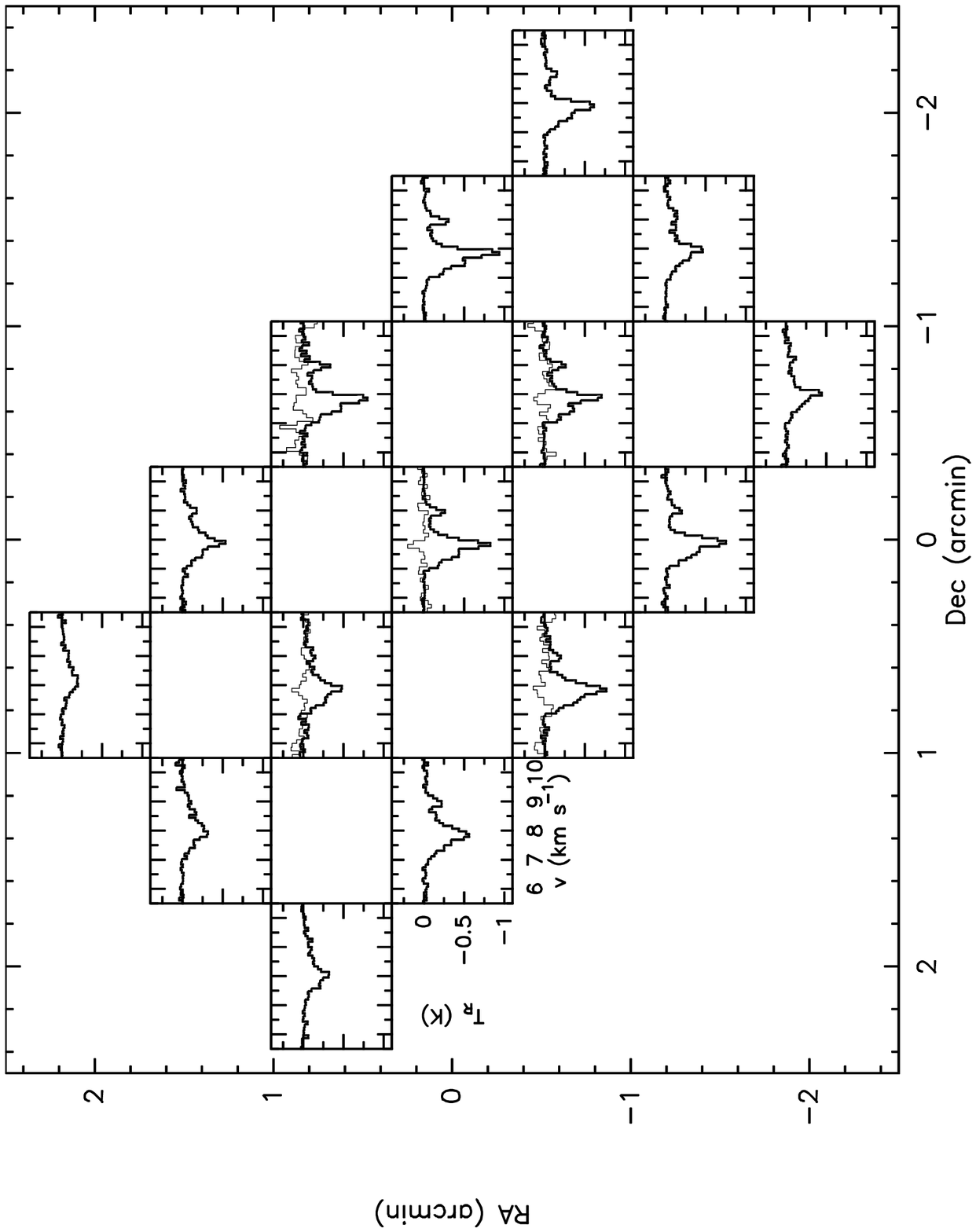} 
\figcaption{\label{l1498map} Map of H$_2$CO 6 cm
absorption (thick lines) and 1.3 mm emission (thin
lines) in L1498. }
\end{figure}

\begin{figure}
\plotone{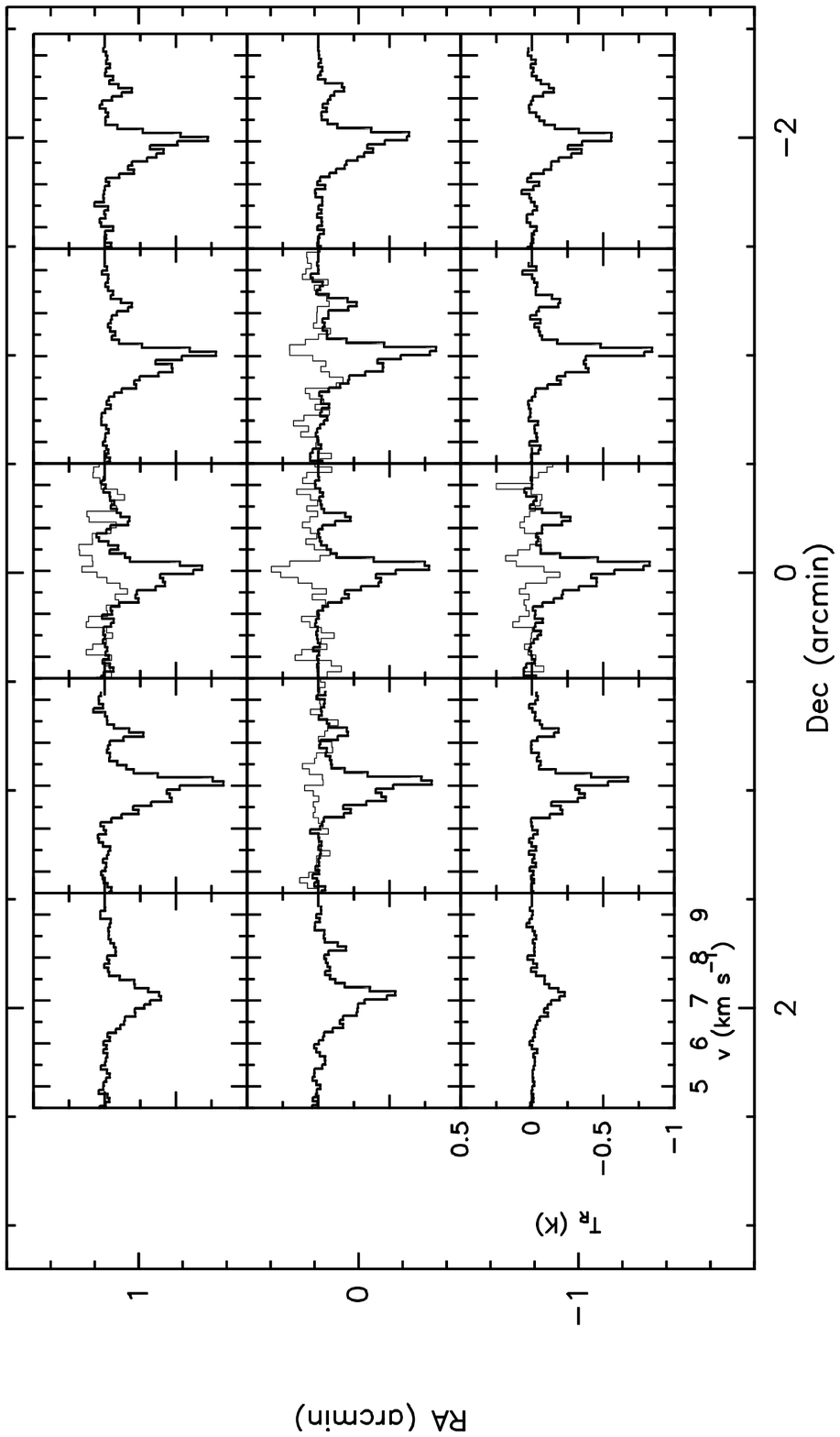}
\figcaption{\label{l1512map} Map of H$_2$CO 6 cm
absorption (thick lines) and 1.3 mm emission (thin
lines) in L1512. }
\end{figure}

\begin{figure}
\plotone{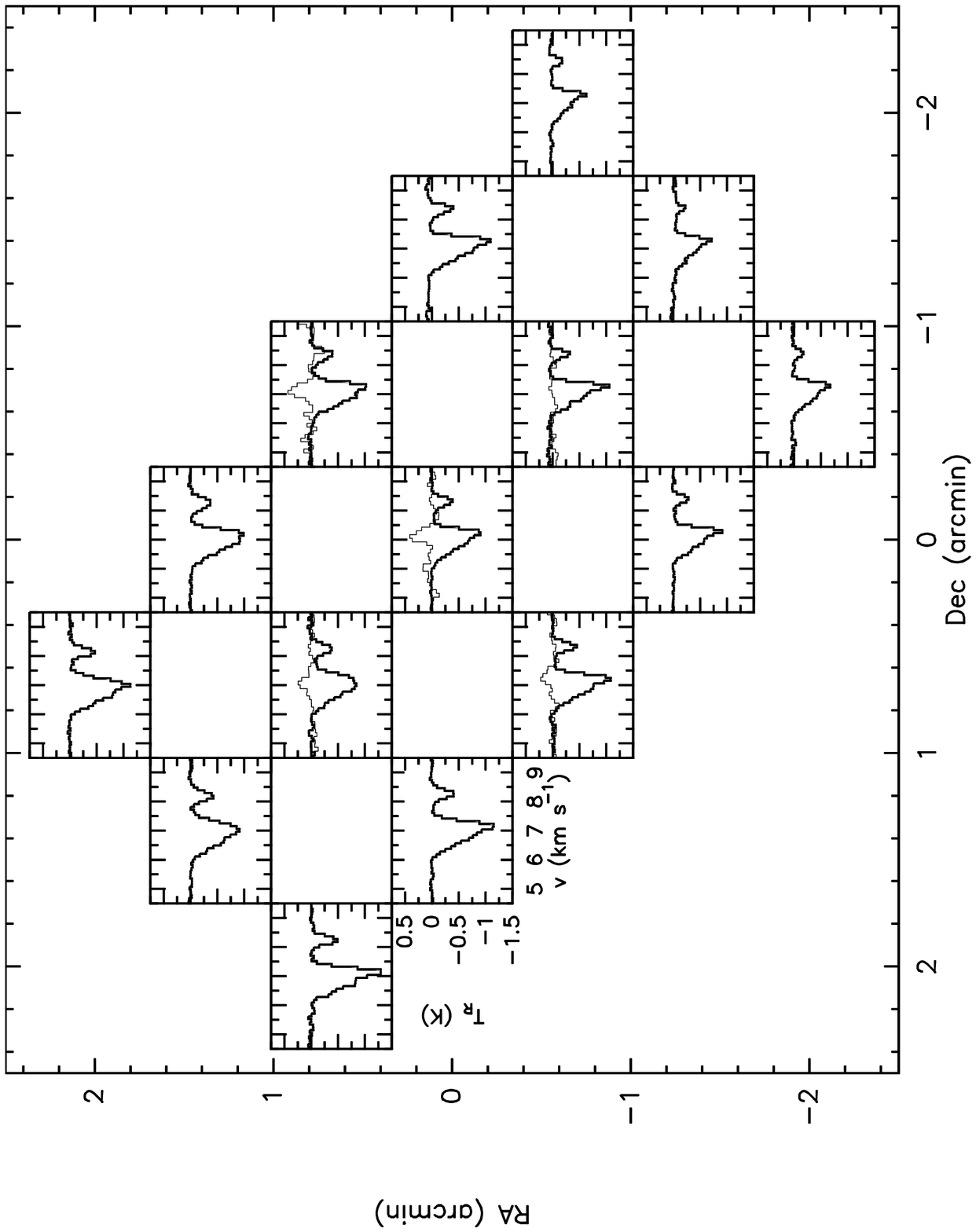} 
\figcaption{\label{l1544map} Map of H$_2$CO 6 cm
absorption (thick lines) and 1.3 mm emission (thin
lines) in L1544. }
\end{figure}

Strong \form\ 6 cm absorption was observed toward every source.  The
absorption spectra obtained at the central positions are shown in
Figure \ref{centers}. The radiation temperature and rms noise at each
observed position is listed in Table \ref{ppctab} and fifteen-point
maps of each PPC are shown in Figures \ref{l1498map} --
\ref{l1544map}. We achieved an average rms of 0.014 K at the center
positions. At the other positions, the average rms was 0.022 K.  The 6
cm line strength stayed nearly constant across all of the PPCs, not
weakening until 120\arcsec\ from the central position.

We have achieved exceptionally good resolution of the hyperfine components
of the 6 cm inversion doublet. The $F=1 - 0$ hyperfine component ($\Delta
\nu$ = $-$18.5 kHz; Tucker, Tomasevich, \& Thaddeus, 1971) was easily
resolved in all of the PPCs.  In L1498 and L1512, the $F = 1 - 2$ and $2-1$
hyperfine components ($\Delta \nu$ = +6.5 and +4.1 kHz respectively)
were also clearly distinguishable from the strong blend of the $F=2 - 2$
and $F=0 - 1$ components, which are closest to the observed central
frequency of 4829.6594 MHz. In L1544, all of the components were blended
except the $F=1 - 0$ component creating an asymmetrical line profile.

The observed 6 cm linewidths are extremely narrow ($\sim$ 0.4 \kms),
indicating low turbulent velocity dispersion. Table \ref{ppctab} lists
the linewidth ($\Delta v$) and the turbulent velocity dispersion
($\sigma_{turb}$) for each source. To calculate the latter, we assumed
$\tk = 10$ K. At the central positions, the \form\ linewidths are
comparable to those observed by Caselli et al.\ (2002a) in the \nthp\
$J=1-0$ line. However, they are about a factor of two smaller than
those observed in the \ammonia\ $(J,K) = (1,1)$ inversion line by
Benson and Myers (1989).

Table \ref{ppctab} also lists the excitation temperature for each
position observed at 6 cm in the PPC maps. The excitation temperature
was calculated using the optical depth and T$_R$ (Heiles 1973). We
assumed that there was only one velocity component and that the
continuum temperature is 2.725 K (Mather et al.\ 1999). We find an
average excitation temperature of 1.5 K. This is consistent with the
results of Heiles (1973), who found a mean excitation temperature of
1.6 K for 9 positions in dark dust clouds using a CBR temperature of
2.8 K.  Vanden Bout et al.\ (1983) found a slightly higher
mean excitation temperature of 1.9 K for 7 dark clouds.

\begin{figure}
\plotone{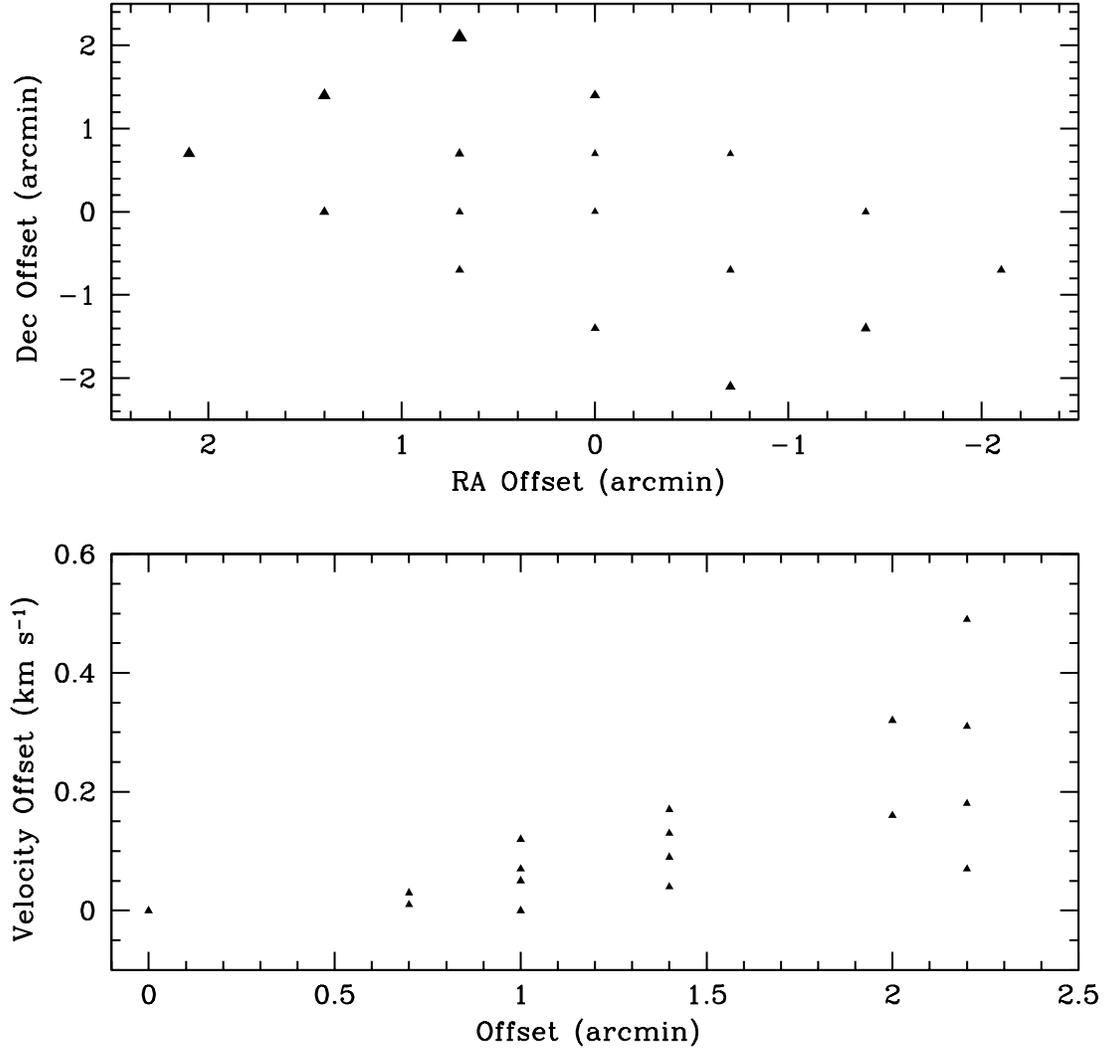} \figcaption{\label{pv} The top panel shows the
velocity gradient in the 6 cm line in L1498 in RA--Dec space.  The
size of the symbol increases with velocity. The central (and smallest)
velocity is 7.85 \kms and the largest velocity is 8.34 \kms. All
velocities are listed in Table \ref{ppctab}. The bottom panel shows
the velocity difference versus the positional offset. This plot shows
that the velocity, as well as the dispersion of velocities, increases
with distance from the center.}
\end{figure}

A significant velocity gradient is seen across L1498. Overall, the
velocity increases with distance away from the center position (Figure
\ref{pv}). This gradient is not consistent with either rotation or
infall, but is suggestive of a cometary or prolate-shaped object seen
face-on. A velocity gradient has been measured in C$^{18}$O (Lemme et
al.\ 1995) and NH$_3$ (Goodman et al.\ 1993). However, the gradient
runs SE--NW, along the main axis of the dust emission, which is very
different from the \form\ gradient.  The apparent velocity gradient
could be the result of two distinct velocity components, one
associated with L1498 at 7.8 \kms\ and a second unrelated component at
a higher velocity. Kuiper, Langer, \& Velusamy (1996) observed a
second component toward L1498 in C$^{18}$O and CS at 8.1 \kms. Tafalla
et al.\ (2002) suggested that the second velocity component might be
the result of gas unrelated to the dark cloud and equivalent to the
``red'' component in CS toward L1498 described by Lemme et al.\
(1995). The 6 cm \form\ observations also show some evidence of a
second velocity component in a small wing near 8.3 \kms\ (see also
Section \ref{l1498_sec} below). Wang (1994) observed the 6 cm \form\
line toward L1498 with a smaller beam at the VLA. Wang did not observe
a velocity gradient, providing further evidence that the velocity
gradient we observed is due to an unrelated second component.

The 1.3 mm observations toward the center of each PPC are shown in
Figure \ref{centers}. The 1.3 mm five-point maps are shown in Figure
\ref{l1498map} -- \ref{l1544map} along with the 6 cm absorption
maps. The average rms for the 1.3 mm observations was 0.057 K. The 1.3
mm line was not detected at one offset position in each
map. Two-$\sigma$ upper limits for $T_R$ are listed in Table
\ref{csotab} for positions with no detection along with $T_R$ and rms
at the other positions for each PPC. The upper limits given are small
in comparison with the detected values, especially in L1512 and L1544.
Therefore, the non-detections may indicate a real decrease of \form\
emission in that direction (southeast in L1498, north in L1512, and
southwest in L1544).  Table \ref{csotab} also lists $v_{LSR}$, $\Delta
v$, and $\sigma_{turb}$ for each position with a detection. The
$\sigma_{turb}$ agree with those found for the 6 cm lines within a
channel width ($\pm$ 0.06 \kms\ for the 1.3 mm data).

\begin{figure}
\plotone{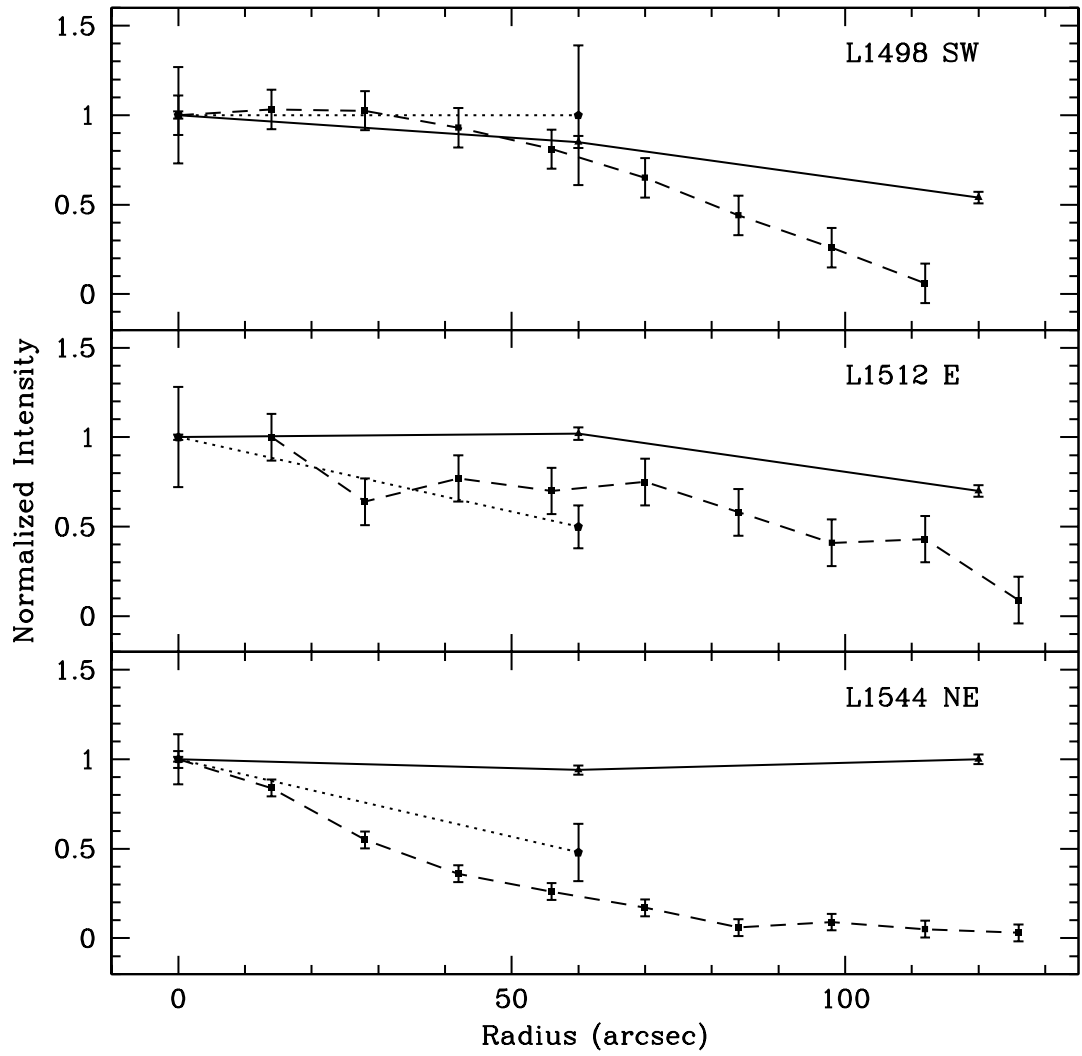} \figcaption{\label{profile} The solid lines
are the normalized radial intensity profiles of the 6 cm \form\ peak
absorption ($T_R$). The dotted lines are profiles of the 1.3 mm \form\
$T_R$. The dashed lines are the profiles of the 850 \micron\ dust
emission.  The slices are taken along the short axis of the PPCs as
seen by the dust in the direction indicated in the figure.  The error
bars represent $\pm$ 1-$\sigma$; the largest error bar at 0\arcsec\ is
for the 1.3 mm data. The center positions are those given in the text
for the \form\ observations. }
\end{figure}

The H$_2$CO 6 cm maps indicate that the \form\ absorption is
considerably more extended than the dust continuum (Evans et al.\
2001; Shirley et al.\ 2000). Vanden Bout et al.\ (1983) found that the
6 cm absorption extended over a region more than 10\arcmin\ in size
toward L1544. However, the 1.3 mm emission, which only extends to
60\arcsec, appears to drop off more quickly than the 6 cm \form\
absorption in L1512 and L1544. Figure \ref{profile} shows normalized
radial profiles of the \form\ $T_R$ for both 6 cm and 1.3 mm and the
850 \micron\ emission along the short axis of the dark clouds. The
broad extent of the 6 cm absorption and the smaller region of 1.3 mm
\form\ emission indicate that the two transitions are probing
different parts of the PPCs. The 1.3 mm emission is likely tracing
denser gas than the 6 cm absorption. This conclusion is discussed
further in the context of our models in Section \ref{sec_dis} below.

\section{Models}\label{sec_mod}

The \form\ line profiles of each PPC were modeled in the final stage
of a three step process. First, dust continuum data for the PPCs were
modeled to determine the density and dust temperature profiles of each
core. Second, the gas temperature structure was determined from the
dust temperature with an energetics code. Finally, the density and gas
temperature profiles were used to model the \form\ lines using a Monte
Carlo method code (Choi et al.\ 1995). The details of each stage are
described below.

The observed sources have dust continuum maps at 450 and 850 \micron\
(Shirley et al.\ 2000). The dust continuum radial intensity profile
for L1498 was modeled by Y.\ Shirley et al.\ (in preparation), and L1512 and
L1544 were modeled by Evans et al.\ (2001) yielding a density and dust
temperature profile for each source.  The temperature structure was
modeled with a one-dimensional radiative transfer code (Egan, Leung \&
Spagna 1988). The density structure was assumed to be that of a
Bonnor-Ebert sphere (Evans et al.\ 2001). The cloud was assumed to
have no internal energy source and to be heated only by the
interstellar radiation field (ISRF). A simulated observation of the
cloud yielded a spectral energy distribution (SED) and a radial
intensity profile. The model SED and radial profile were compared to
the observations. The physical parameters of the model cloud were
iteratively adjusted until a good fit was found to the
observations. The best-fit model was judged by the reduced chi-squared
of the model fit to the 450 and 850 \micron\ radial profiles and the
SED.

\begin{figure}
\plotone{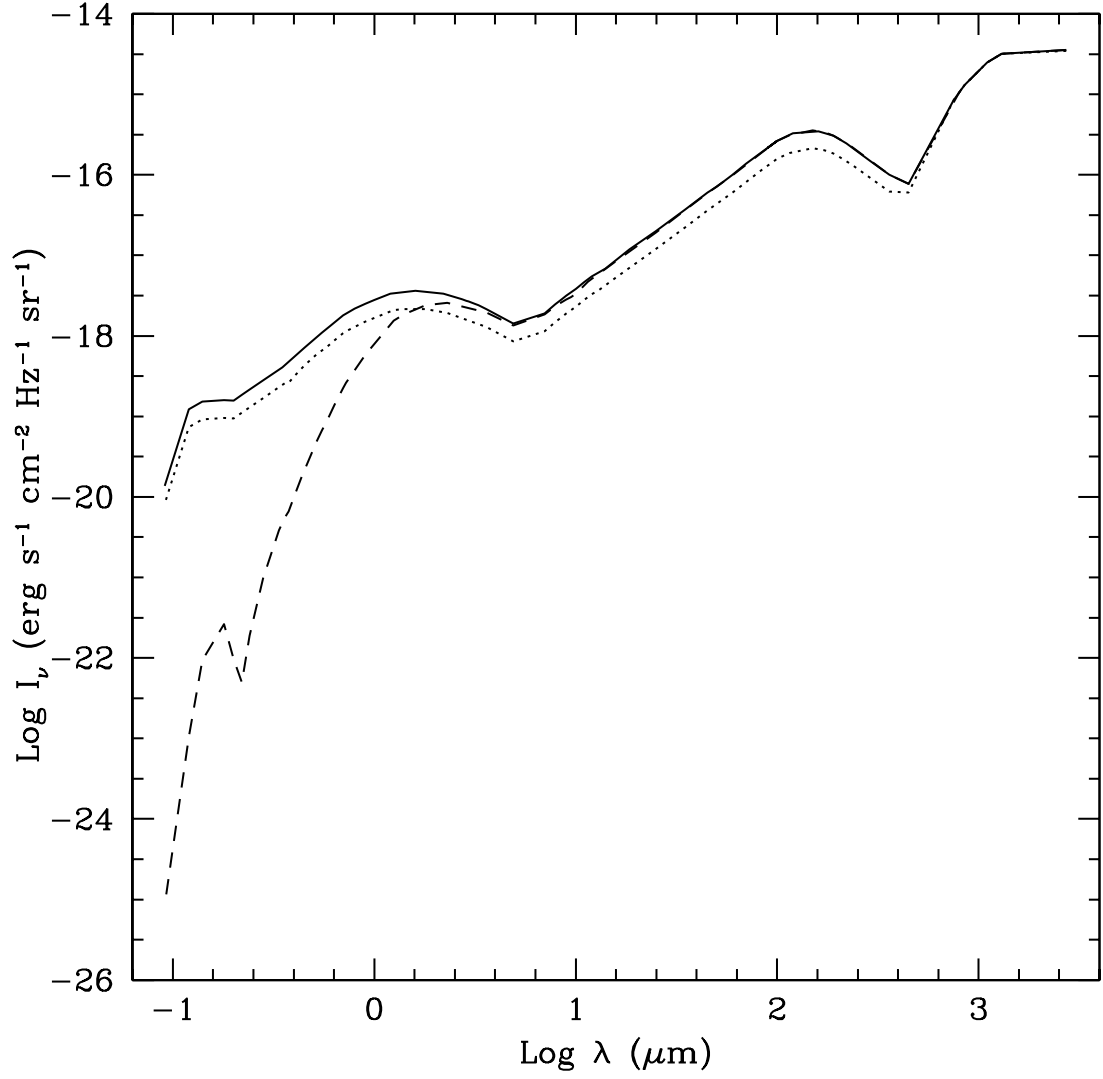} \figcaption{\label{isrf} Plot of the standard ISRF
(solid line), the reduced ISRF (dotted line), and the attenuated ISRF
(dashed line). The reduced ISRF is weaker than the standard ISRF
everywhere by a factor of 1.66. The attenuated ISRF ($A_V$ = 3) is
reduced from the standard ISRF predominately in the visible and UV. }
\end{figure}

We remodeled L1512 and L1544 with a slightly different method than
that used by Evans et al.\ (2001) but in accordance with Y.\ Shirley
et al.\ (in preparation). Our models differed from the Evans et
al. (2001) models in how the ISRF was treated.  Evans et al. (2001)
found that an ISRF reduced from the Black-Draine ISRF, which they
adopted as a standard, improved the fit to the dust emission for L1512
and L1544. In the model for L1512, the ISRF (except for the CBR
component) was multiplied by a factor of 0.3 and, in L1544, by a
factor of 0.6. In the models presented here, the ISRF was instead
attenuated to correspond to a certain visual extinction ($A_V$). The
attenuated ISRF is reduced much more at ultraviolet and visible
wavelengths but much less in the infrared than the ISRF reduced by a
single muliplicative factor. Figure \ref{isrf} plots both the reduced
and attenuated ISRF for comparison.

The $A_V$ values resulting from the dust models correspond well to the
known environments of the PPCs. The best-fit dust model of L1498
required two magnitudes of visual extinction surrounding the core (Y.\
Shirley et al., in preparation). Cambr\'{e}sy's (1999) extinction map
of Taurus shows L1498 to be fairly isolated and surrounded by ambient
gas with $A_V = 1-2$ magnitudes.  L1512 is on the edge of Taurus where
$A_V$ is about 0.5 magnitudes (Falgarone et al.\ 1998). For this PPC,
attenuating the ISRF did not improve the fit to the dust, but since it
was not a significantly worse fit, the $A_V=0.5$ attenuated dust model
was used for consistency. We found that attenuating the ISRF in the
models for L1544 fit the dust emission radial profiles better than
simply reducing the ISRF. L1544 is the most heavily embedded core; our
dust radiative transfer model indicates $A_V = 3$. This value is
consistent with the extinction determined from star counts at a
5\arcmin\ radius ($A_V=2-3$, Minn 1991; Snell 1981).

Once the dust temperature profile had been determined from the dust
emission, an energetics code (after Doty \& Neufeld 1997) was used to
calculate the temperature of the gas from the temperature of the
dust. The calculation accounted for radiative cooling of the gas,
transfer of energy from collisions between the gas and dust grains,
heating by cosmic rays, and photoelectric heating. The Appendix
provides complete details of the energetics calculations.

The strength of the heating on the outside of the cloud in the
energetics code is primarily controlled by the factor $G_0$, which is
the strength of the UV field relative to the standard ISRF at the
outer PPC boundary. A value of $G_0$ less than unity results from the
attenuation of the ISRF by the material surrounding the PPC, and a
smaller $G_0$ indicates greater attenuation.  In order to constrain
the estimate of $G_0$ for each cloud, we modeled the CO $J=1-0$ line
and compared the models to observations for each PPC.  This CO
transition is completely thermalized. By comparing the strength of the
modeled and observed CO lines, we are able to determine if the model
temperature is too hot or cold on the outside of the cloud and,
therefore, if $G_0$ is too high or low.  We used the CO observations
of Kuiper et al.\ (1996), Falgarone et al.\ (1998), and
Tafalla et al.\ (1998) for L1498, L1512, and L1544 respectively. We
did not attempt to fit the CO line profile in detail, because our
models have many parameters other than $G_0$.  The CO models all had
the same abundance and physical parameters. Different $G_0$ values
were tested to find the one that produced a $T_R$ that was consistent
with the observations.

We compared $G_0$ found using CO observations as a constraint to $G_0$
calculated from the $A_V$ value used to attenuate the ISRF in the dust
model (see the equation for $G(r)$ in the Appendix) for each
source. The CO-derived $G_0$ values were 0.06, 0.03, and 0.08 for
L1498, L1512, and L1544. The $G_0$ values calculated from the dust
model $A_V$ were 0.03, 0.005, and 0.41, respectively. The $G_0$ values
derived from CO are larger than those calculated from the $A_V$ values
for L1498 and L1544, indicating less extinction toward the PPCs. For
L1512, the $G_0$ derived from CO is much smaller than the value
calculated from the dust model $A_V$ but is consistent with the $G_0$
values derived from CO for the other sources. Complete
photon-dominated region (PDR) analysis is needed in order to fully
examine this inconsistency. The CO-derived $G_0$ values are used in
the energetics code for this work, because the CO is much more
sensitive to $G_0$ than the dust models are to $A_V$. Further, the
equation used to calculate $G_0$ for a given $A_V$ is for
plane-parallel slabs and is not appropriate for spheres as discussed
in the Appendix.

We used the best fit density profile from the dust model and the gas
temperature profile from the energetics calculations as input into a
Monte Carlo (MC) method radiative transfer code (Choi et al.\ 1995) to
model the H$_2$CO 6 cm and 1.3 mm lines. Figure \ref{temp_den} shows
the density and temperature profiles resulting from the dust and
energetics models for each PPC. The results from the MC code were
convolved to the resolution of the observations for comparison with
the data.  The MC model has two free parameters, the outer radius of
the cloud ($R_{out}$) and the microturbulent velocity dispersion
($\delta v_t$). The same outer radius was used for the dust and MC
models (0.15 pc) for consistency; $\delta v_t$ was iterated on to find
the best fitting value for each PPC. Theoretical collision rates for
\form\ only exist down to 10 K, and the temperatures in dark starless
cores drop below that. Therefore, for the purpose of these models, the
\form\ collision rates were linearly extrapolated down to 3 K and 5 K
from the known rates at 10 K and 15 K (Green 1991). The convergence
criterion in the MC code was set at 5\%. A more stringent convergence
criterion of 2\% was tested and did not significantly affect the model
results.

The components of the hyperfine structure of the 6 cm line were
modeled independently with the exception of the $F=2-2$ and $F=0-1$
components. These two components lie within 0.1 \kms\ of one another,
less than the turbulent velocity dispersion in the observed
lines. Therefore, the relative intensities of these lines were added
together for the purpose of our models. The frequency of the combined
component was given as the relative-intensity-weighted mean of the
individual frequencies. The five modeled components were added for the
complete modeled 6 cm line profile.  We compared this method of
modeling five components to modeling the full six hyperfine components
and found that using all six components deepened the absorption by
only about 10\%.

Our models indicate that H$_2$CO is significantly depleted in these
PPCs as predicted by various chemical models (Lee et al.\ 2004; Y.\
Aikawa 2003, private communication).  Figure \ref{model} shows the
simulated 6 cm line profile for a dark cloud with a central density of
10$^6$ cm$^{-3}$ and an undepleted H$_2$CO fractional abundance of
10$^{-8}$. The model is plotted with the observed line profile of
L1544 to show the large discrepancy between the expected undepleted
line profile and what was actually observed. The undepleted model
predicts 6 cm {\it emission} toward the center of the PPC, a {\it
qualitative} discrepancy with the observation. Depletion significantly
affects how the 6 cm \form\ transition probes cold, dense gas. 
If \form\ is depleted in the center of a PPC, then the 6 cm line will not
be able to see the densest part of the core. 

Depletion was introduced into the models with an abundance profile
characterized by less \form\ at small radii than at larger
radii. Several different functional forms of the abundance profile
were compared in the modeling process, including a power law,
exponential, and step function (see Table 6 in Lee et al.\ 2003).  The
abundance at the outer edge of the cloud was set by $X_0$, the
undepleted abundance. The shape of the step function and exponential
were characterized by a radial scale of depletion ($r_D$). The depth
of the step function was set by ($f_D$), the factor by which the
abundance in the central portion of the PPC was decreased.

These simple functional forms of the abundance profile were also
compared with the profile generated by chemical network models (Lee et
al.\ 2004), where the reactions between gas and grains, such as
depletion and desorption of species, as well as gas phase chemistry
have been considered.  Lee et al.\ (2004) used a series of
Bonnor-Ebert spheres of various central densities ($n_c$) to calculate
the evolution of chemistry in the PPC stage.  They assumed that a
Bonnor-Ebert sphere evolves from lower central density to higher
central density, and the timescale of the evolution decreases as the
central density increases.  The total timescale for the PPC stage was
assumed to be a million years.  The abundance profile resulting from
the chemical models was multiplied by a scale factor ranging from 0.3
to 2.0 in the models to best fit the data since the actual timescale
is unknown. Figure \ref{abun} shows examples of the various abundance
profiles. The values specifying the exact shape of the abundance
profiles ($X_0$, $r_D$, $f_D$, $n_c$, etc.) were iterated on until a
best fit for each type of profile was found for each source. The
results of the models with different abundance profiles are compared
below (Sections \ref{l1498_sec}--\ref{sec_dis} and Table
\ref{modtab}).

Each model was compared to the observed 6 cm line at the center of
each PPC and at offsets of 60\arcsec\ and 120\arcsec\ from the center.
The model was also compared to the 1.3 mm line at the center and
60\arcsec\ offset. Figures \ref{l1498mod} -- \ref{l1544mod} plot the
data and the best fitting model for each source. The offset spectra
are the weighted average of the spectra in all directions at either
60\arcsec\ or 120\arcsec\ from the center of the cloud.  For the 6 cm
line, the 60\arcsec\ offset spectrum is the average of 8 spectra, and
the 120\arcsec\ spectrum is the average of 6 spectra for each
source. For the 1.3 mm line, the 60\arcsec\ offset spectrum is the
average of 4 spectra.

The best model was selected by comparing values of the absolute
deviation that characterize the model fit to the data at both
wavelengths and all offsets. The absolute deviation is defined as the
sum of $|\mathrm{Model}(i) - \mathrm{Observed}(i)|$ over all data
points $i$ (Press et al.\ 1992). The absolute deviation was chosen
instead of the more common reduced chi-squared value in order to
eliminate artificial weighting of the noisier 1.3 mm data. For nearly
all the models, the 1.3 mm reduced chi-square was low compared to that
for the 6 cm lines even when the model clearly fit the 6 cm data
better than the 1.3 mm data. The high noise levels in the 1.3 mm data
dramatically decreased the reduced chi-square values, and, as a
result, created a bias toward the 1.3 mm fits. The absolute deviation
does not consider noise so it is free of this bias.

We have calculated the absolute deviation in two different
ways. First, we calculated the absolute deviation between the
integrated intensity of the model and the observations. A second
absolute deviation compares the line profile shape of the model and
data. Each of these values is calculated for 6 cm and 1.3 mm lines
independently. The absolute deviation is also calculated independently
for each observed offset. The average absolute deviation over all
offsets for each wavelength is reported in Table \ref{modtab}.

We chose the model with the lowest integrated intensity and profile
absolute deviation values at 6 cm for the best-fit model because the 6
cm data have high signal-to-noise and provide good constraints for the
models. All of the models presented in Table \ref{modtab} are the
best-fitting models for particular abundance profiles. There was often
a discrepancy between the models with the lowest absolute deviation
for the 6 cm data and the 1.3 mm data. We found that no reasonable
models for the 6 cm data fit the 1.3 mm data well. This discrepancy is
discussed in further detail in Section \ref{sec_dis} below. The models
are discussed by source in Sections \ref{l1498_sec} --
\ref{l1544_sec}.

\subsection{L1498}\label{l1498_sec}

L1498 is the most diffuse of the three PPCs discussed here. As a
result, the best-fitting dust model has the lowest central density,
\eten{4} \cmv, of all the sources (Figure \ref{temp_den}). L1498 was
modeled with an outer radius of 0.15 pc. This is smaller than the
outer radius of the best-fitting dust radiative transfer model (0.17
pc, Y.\ Shirley et al., in preparation) but is consistent with the outer
radii used by Lee et al.\ (2004) for the chemical models and for L1512
and L1544 in this work. All the L1498 models use $\delta v_t$ = 0.1
\kms. The $G_0$ derived from the CO observations of L1498 is
0.06. Table \ref{modtab} gives a full listing of model parameters and
absolute deviations for models of L1498 with differing abundance
profiles.

The best-fitting model for L1498 (Figure \ref{l1498mod}) uses a
chemical model abundance profile (Figure \ref{abun}). The chemical
model abundance profile was multiplied by a factor of 0.8 in order to
best fit the data. The chemical model used had the same $A_V$ as the
dust model for L1498, $A_V=2$, but a much higher central density of
\eten{7} \cmv. The higher central density in the chemical model
results in more central depletion of \form\ in the abundance
profile. Further, the chemical model abundance profile used to fit the
L1498 data is dramatically different from the other abundance profiles
at large radii (Figure \ref{abun}). The chemical model abundance
profile has a high maximum abundance ($\sim$ 9\ee{-9}) and decreases
rapidly at large radii for $A_V=2$. This abundance profile resulted in
a model that fit the strength of the the 6 cm \form\ absorption well
at 60\arcsec\ and 120\arcsec. The absorption predicted by the model at
the center position appears too weak. The weak central line could
indicate that there is not enough depletion at the center of the
model, and the main hyperfine component is being driven toward
emission. Alternatively, there could be less absorption in the main
component because it is too optically thick in the model and is not
probing as far into the cloud as the other components. Abundance
profiles which have more central depletion but no outer decrease, fit
the central 6 cm line better but were too deep at the offset
positions.

An alternative model for L1498 was found by allowing $G_0$ to be a
free parameter rather than constraining it with CO observations. An
abundance profile, adopted from the chemical model with $A_V$=2 at the
time step at which $\rm n_c=10^4$ cm$^{-3}$ that is consistent with
the model of the dust emission by Y.\ Shirley et al.\ (in preparation), can
fit the observed 6 cm \form\ profiles if a $G_0$ 10 times higher than
what was found from the CO $J=1-0$ line is used. The absolute
deviations for this model are listed in Table \ref{modtab} along with
the lower $G_0$ models. This model reproduces the 6 cm line profiles
at all offsets well and does not have the same problem of weak center
absorption as described above. In this chemical model, \form\ is not
very depleted even at the center because the depletion timescale is
long at low density. This result shows that the strength of the ISRF
($G_0$) can greatly affect line profiles as discussed in Lee et
al.\ (2004). Full PDR analysis and a better understanding of the
environment is needed to correctly determine $G_0$. Therefore, for the
purposes of consistency in this paper, we have constrained $G_0$ by CO
$J=1-0$ observations with the caveat that the $G_0$ values derived may
be underestimated.

The best-fitting model of L1498 illustrates a problem common in all
our models.  We did not adequately fit both the 6 cm data and the 1.3
mm data with the same model parameters. The 1.3 mm models are
consistently weaker than the observed emission lines when the 6 cm
data is fitted well. For L1498, the best-fitting chemical model
abundance profile did result in a better fit to the 1.3 mm data than
the other abundance profiles modeled, although it was still too weak
(Figure \ref{l1498mod}). The better fit to the 1.3 mm data may have
resulted from the fact that the chemical model abundance profile has a
higher maximum abundance than the other profiles, although it drops
off to comparatively much lower abundances at large radii (Figure
\ref{abun}). In general, the 1.3 mm data could be fit with a larger
$X_0$, which results in too much absorption at 6 cm. The difficulty in
fitting both the 6 cm and 1.3 mm data with one model is discussed
further in Section \ref{sec_dis} below.

The model also does not fit the data well to the right of the main 6
cm \form\ hyperfine component in L1498. The data show a wing around
8.3 \kms\ that the models do not reproduce. We postulate that this
wing is due to a second velocity component caused by unrelated gas. A
second velocity component has also been observed in C$^{18}$O and CS
(Kuiper et al.\ 1996) as described above in Section \ref{sec_res}.

\subsection{L1512}

L1512 is best fitted with a model of a moderate central density ($n_c
= 10^5$ \cmv) and a step function abundance profile.  All the L1512
models had an outer radius of 0.15 pc, a $\delta v_t$ of 0.1 \kms, and
$G_0$ = 0.03. Model parameters and results are given in Table
\ref{modtab} and Figures \ref{temp_den} and \ref{l1512mod}. As Figure
\ref{l1512mod} shows, the step function model predicts less 6 cm
absorption in the main hyperfine component ($F=2-2$) than observed at
the center of L1512. The other abundance profiles modeled predict even
less absorption than the step function model. The step function
abundance profile differs significantly from the power law,
exponential, and chemical model profiles in that \form\ is depleted
out to a larger radius. The other profiles fall off gradually, but
the step function has a steep drop in abundance at 0.05 pc. Although
the best-fit model is not a perfect match to the data, the model
constrains the \form\ abundance profile by ruling out profiles that
are depleted too slowly toward the center of the core. The parameters
characterizing the best-fit step function ($r_D$, $f_D$, $X_0$) are
good within a factor of two to fit the data.

Again, the best-fit model produces 1.3 mm line profiles that are too
weak compared to the data (see Section \ref{sec_dis} below).
Additionally, there is a velocity shift between the model and 1.3 mm
data at the 60\arcsec\ offset position. The central velocity of the
data appears higher than that of the model.  The map of L1512 in
Figure \ref{l1512map} shows that there is no obvious velocity gradient
in the 6 cm data as is the case in L1498.

As with L1498, the models do not fit the wing to the right of the main
6 cm hyperfine component as shown in Figure \ref{l1512mod}, and we
believe that this again is the result of a second velocity
component. Falgarone et al.\ (1998) observed multiple velocity
components in several transitions of CO toward L1512. They observed a
second velocity component in the line profile of C$^{18}$O at a
similar velocity (7.3 -- 8.1 km s$^{-1}$) to the wing in the H$_2$CO
line.

\subsection{L1544}\label{l1544_sec}

\begin{figure}
\plotone{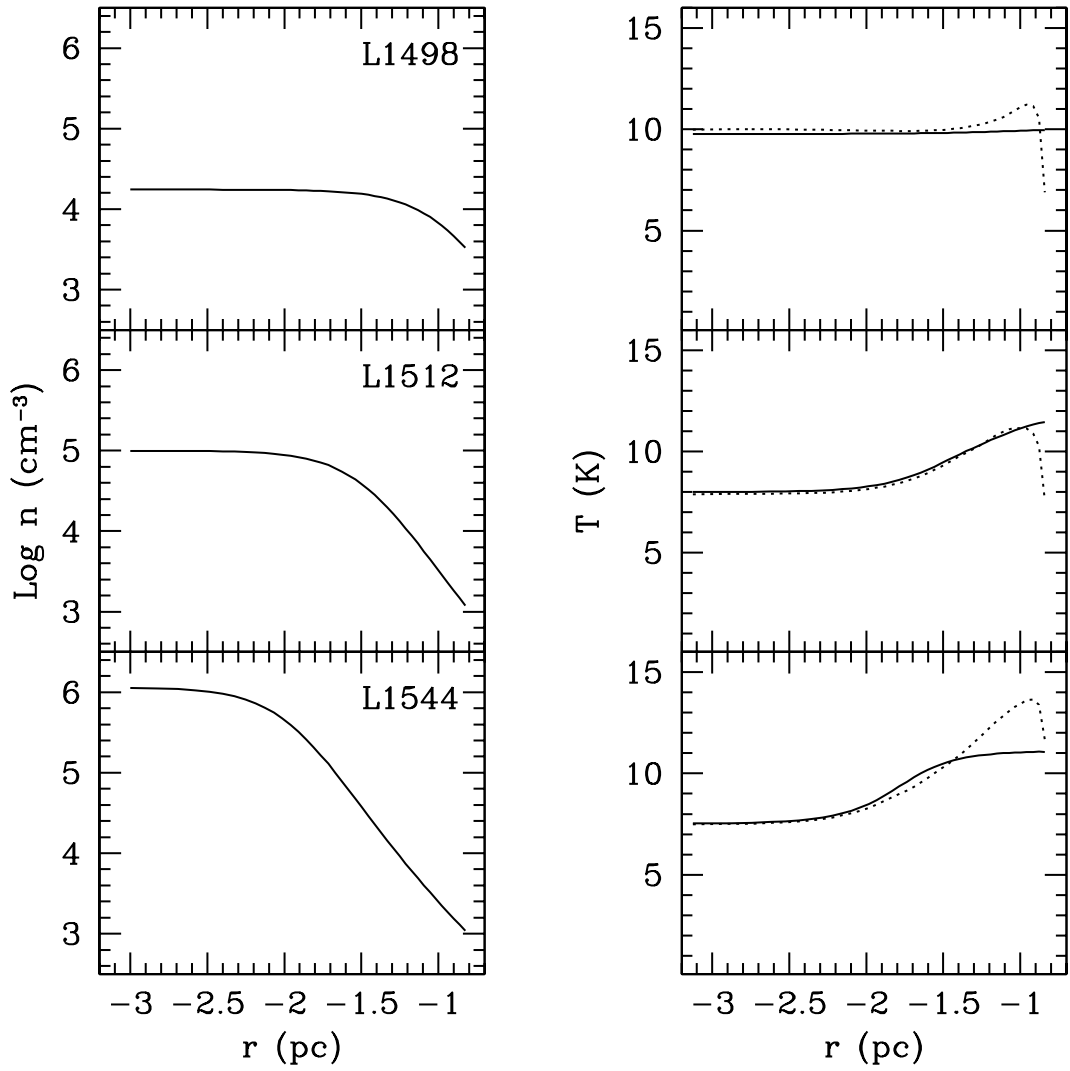} 
\figcaption{\label{temp_den}Model density and temperature versus radius for 
the three PPCs studied here. 
In the left column, the curve shows the density profile.  
In the right column, the solid line is the dust temperature
and the dotted line is the gas temperature.}
\end{figure}

\begin{figure}
\plotone{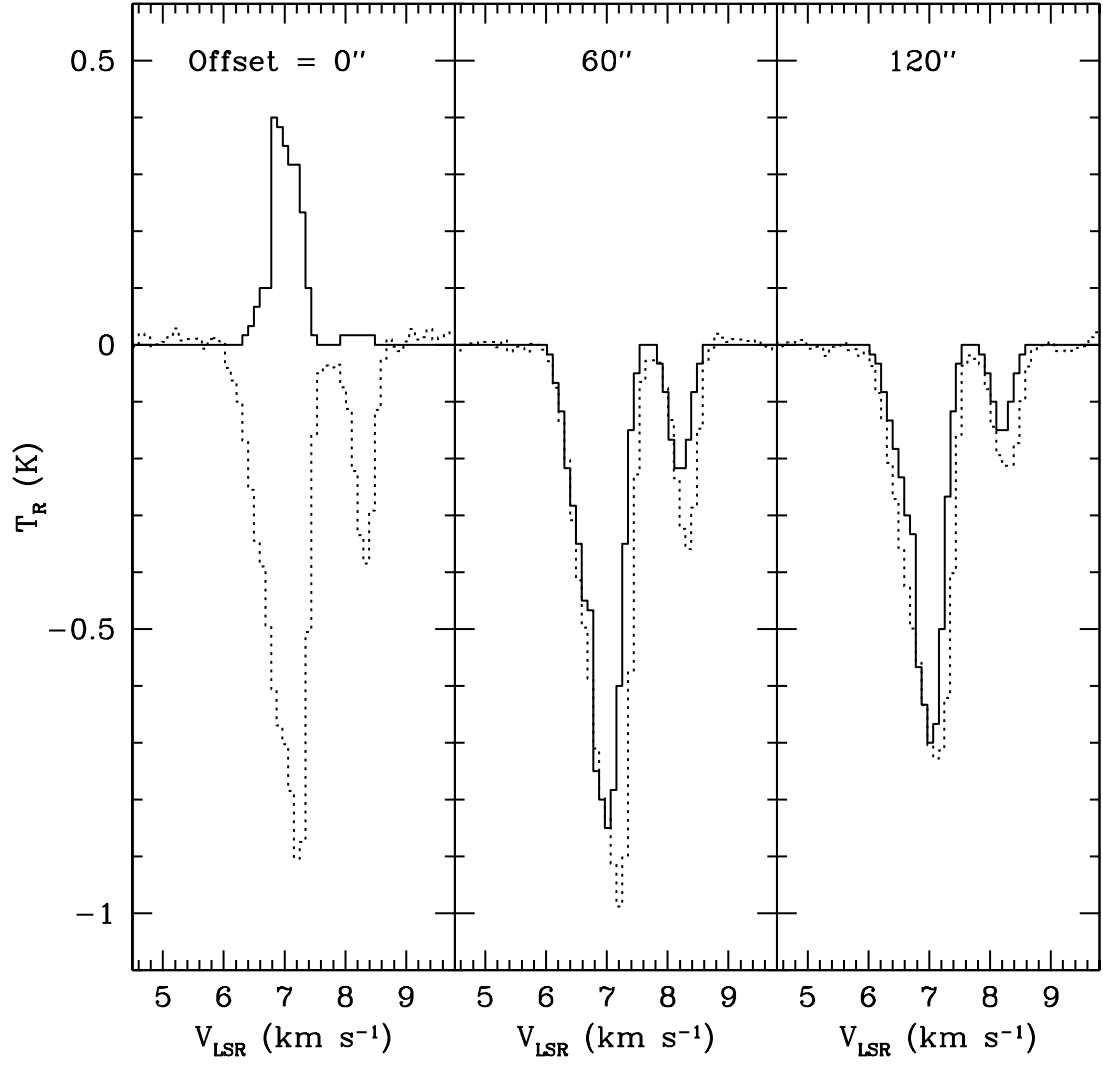} \figcaption{\label{model} The solid line is
a model of the expected H$_2$CO 6 cm line profile in a cold dense
cloud with a central density of 10$^6$ cm$^{-3}$ and H$_2$CO abundance
of 10$^{-8}$. The dotted line is the observed spectrum of the 6 cm
line in L1544. The figure illustrates that \form\ must be depleted
toward the center of the PPC.}
\end{figure}

\begin{figure}
\plotone{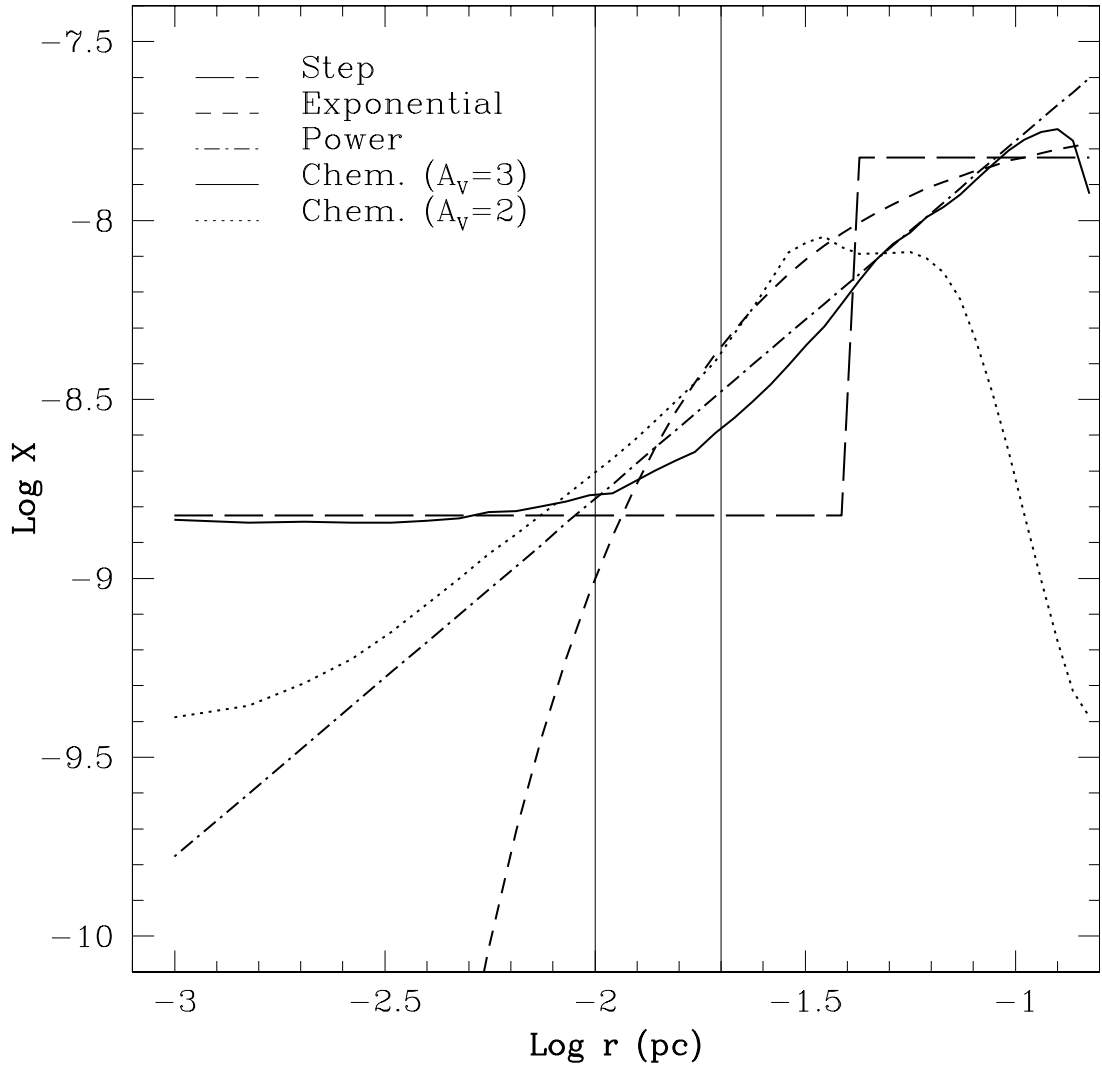} \figcaption{\label{abun} Examples of abundance
profiles used in the models. The long dashed line is a step
function. The short dashed line is an exponential. The dot-dash line
is a power law. The solid line is the result of a chemical model with
$A_V=3$.  The model parameters for these profiles are the same as for
L1544 and are given in Table \ref{modtab}. The dotted line is the
chemical model abundance profile for L1498 with $A_V=2$. The amount of
external extinction dramatically affects the chemical model abundance
profile at large radii. The two vertical solid lines mark the radius
of the beam for the 1.3 mm and 6 cm observations, 15\arcsec\ and
30\arcsec, assuming a distance of 140 pc.}
\end{figure}

\begin{figure}
\plotone{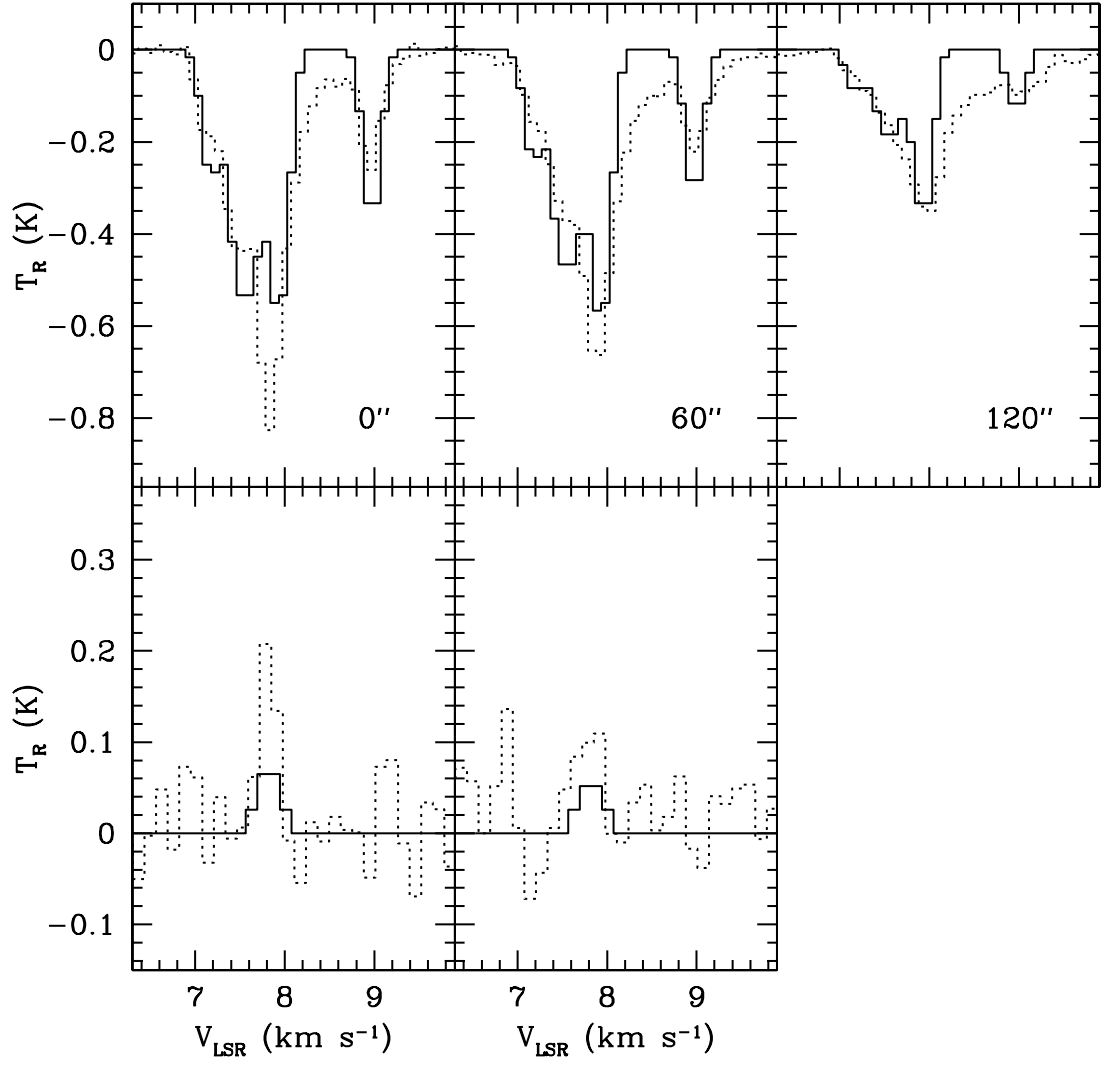} \figcaption{\label{l1498mod} The dotted lines
are the H$_2$CO data, and the solid lines are the best-fit MC model
for L1498. The top panels show the 6 cm line at the offsets indicated
in the figure, and the bottom panels are the 1.3 mm line. The model
has a chemical model abundance distribution. Model parameters are
given in Table \ref{modtab}.}
\end{figure}

\begin{figure}
\plotone{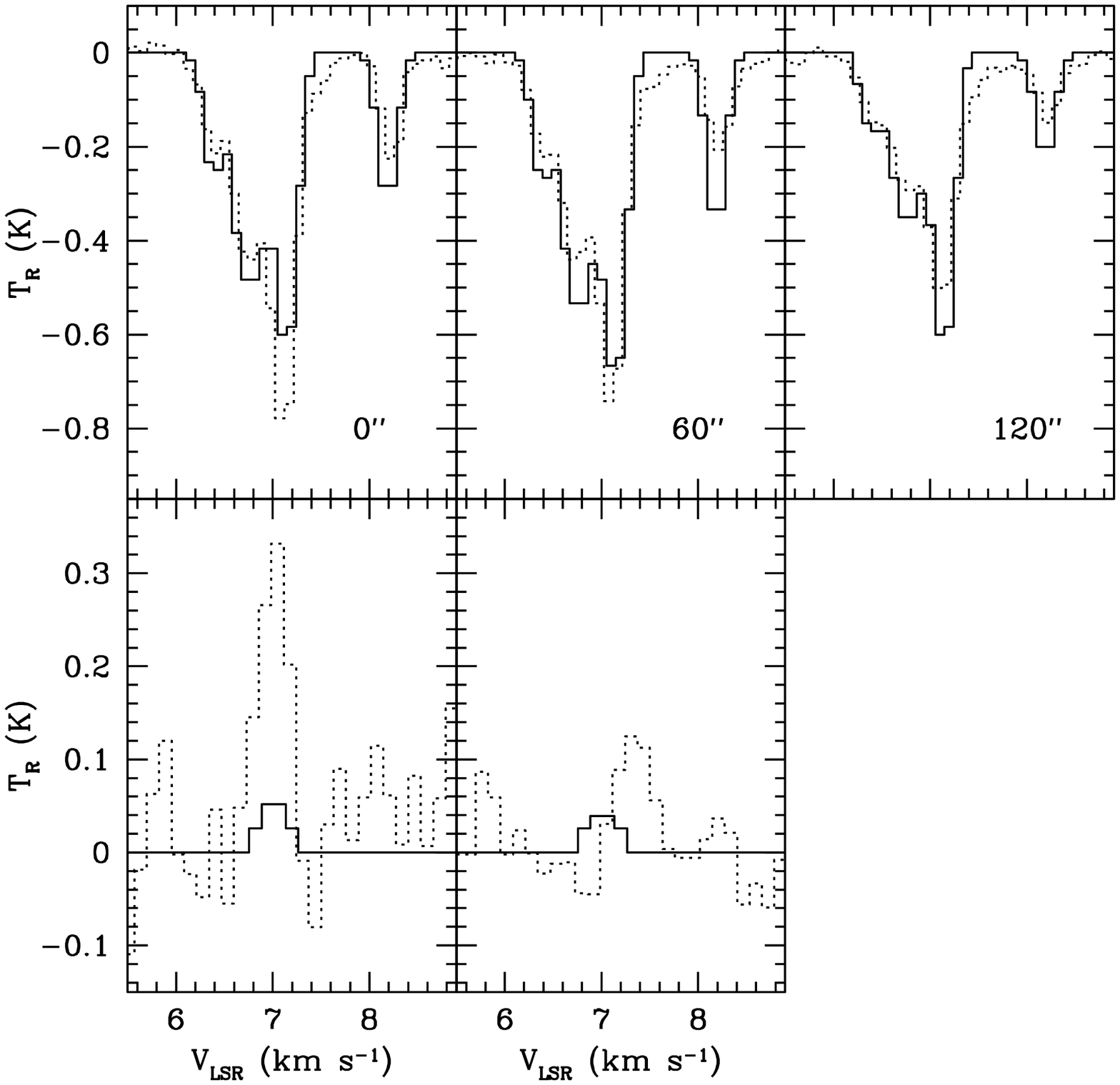} \figcaption{\label{l1512mod} The dotted lines
are the H$_2$CO data, and the solid lines are the best-fit MC model
for L1512.  The top panels show the 6 cm line at the offsets indicated
in the figure, and the bottom panels are the 1.3 mm line. The model
has a step function abundance distribution. Model
parameters are given in Table \ref{modtab}.}
\end{figure}

\begin{figure}
\plotone{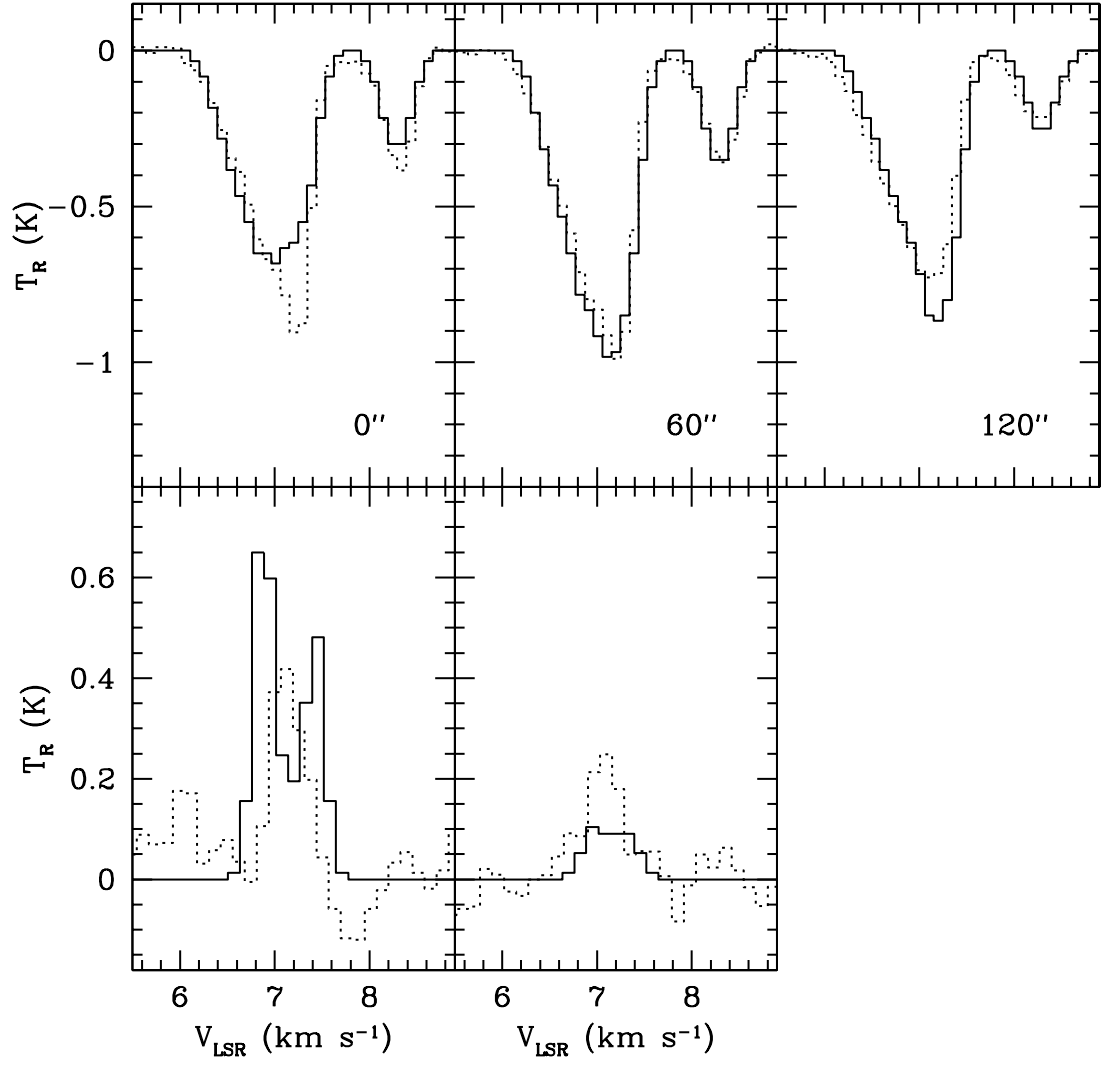} \figcaption{\label{l1544mod} The dotted lines
are the H$_2$CO data, and the solid lines are the best-fit MC model
for L1544.  The top panels show the 6 cm line at the offsets indicated
in the figure, and the bottom panels are the 1.3 mm line. The model
has a power law abundance distribution and a Plummer-like velocity
profile. Model parameters are given in Table \ref{modtab}.}
\end{figure}

L1544 is the most centrally condensed of the three PPCs. The models
require a high central density ($n_c=\eten{6}$ \cmv; Figure
\ref{temp_den}) and a larger $G_0$ of 0.08 compared to the other PPCs.
All the L1544 models have an outer radius of 0.15 pc and $\delta v_t$
of 0.19 \kms. There was no clear best-fitting model for L1544. The
absolute deviation values were very similar for models with four
different abundance profiles. The model parameters and absolute
deviations are shown in Table \ref{modtab}.  Even though the exact
shape of the abundance profile for L1544 cannot be determined by these
models, the general shape of the abundance profile is constrained. All
of the best-fitting abundance profiles have a similar shape at large
radii, are depleted toward the center on a radial scale ($r_D$) of
0.03 to 0.04 pc, and have a outer edge undepleted \form\ fractional
abundance ($X_0$) of about 2$\times 10^{-8}$ (Figure
\ref{abun}). Changes in $r_D$ and $X_0$ of 25\% produce significantly
different model results and absolute deviations, showing that the
reported values characterize the general shape of the \form\ abundance
profile.  Figure \ref{l1544mod} shows the model fit to the data for a
power law abundance profile. The power law models have slightly lower
absolute deviations for the 6 cm data than the models with different
abundance profiles.

In the L1544 models, the model 1.3 mm line at the center position
shows self-absorption (Figure \ref{l1544mod}).  Self-absorption is not
seen in our observations, although the spectra are noisy. We have
tried to reconcile our models and the data in two ways. First, the
density at large radii representing the ambient density of a molecular
cloud was increased from 1000 \cmv\ to 3000 \cmv.  The modeled 1.3 mm
lines were stronger but still showed deep self-absorption in the
models with larger outer density. L1544 is thought to be undergoing
infall (Tafalla et al.\ 1998; Caselli et al.\ 2002b), and we have
incorporated an infall velocity into the model. Two different velocity
profiles were used, an ambipolar diffusion model from Ciolek \& Basu
(2000, time $t_3$ in Figure 2) and a Plummer-like model (Whitworth \&
Ward-Thompson 2001; Lee et al.\ 2003 (Fig.\ 4)).  The absolute
deviation values for a model with a power law abundance profile and a
Plummer-like velocity profile are listed in Table \ref{modtab}. For
both types of velocity profiles, the effects on the model results were
small for the 6 cm lines. The 1.3 mm modeled lines still retained the
self-absorption feature. However, the 1.3 mm lines were also
asymmetrical, skewed to the blue, at both offsets (Figure
\ref{l1544mod}). This effect was more pronounced with the Ciolek \&
Basu (2000) velocity profile. The noise makes the determination of an
asymmetrical profile difficult in the 1.3 mm data for L1544. A skewed
1.3 mm line profile cannot be ruled out.

\subsection{Discussion}\label{sec_dis}

An unambiguous aspect of our models is that all require depletion of
formaldehyde in the center of the core. Depletion of carbon-bearing
molecules such as CO and CS in the center of PPCs, including L1498,
L1512 and L1544, is well known (e.g., Tafalla et al.\ 2004; Lee et
al.\ 2003, Tafalla et al.\ 2002). Tafalla et al.\ (2004) modeled high
angular resolution C$^{18}$O and CS observations toward L1498 with a
similar method to the one in this work, except that they assumed a
constant dust temperature and used a larger outer radius (0.25
pc). They found the high resolution data required a faster decrease in
abundance than their previous models of lower resolution data had
suggested (Tafalla et al.\ 2002). Their result is consistent with the
findings presented here that, although the exact shape of the
abundance profile cannot be uniquely determined, all the models
suggest a significant drop in abundance toward the center of PPCs.
Tafalla et al.\ (2004) fit the CO and CS data with a step function
abundance profile that drops to zero at a given radius. They found
depletion radii about a factor of two larger for CO and CS than the
depletion radius used in the step function abundance profile model of
\form\ in L1498 (Table \ref{modtab}). Lee et al.\ (2003) also found
C$^{18}$O to be depleted out to larger radii ($r_D$ = 0.075 and
0.045 pc) and by a larger factor ($f_D$=25) in L1512 and L1544 than is
required for \form\ in the models presented here.

The depletion of H$_2$CO in PPCs may be related to the chemical
processes of deuteration and the reaction between gas and
grains. H$_2$CO is likely depleted in PPCs due to accretion onto cold
dust grains (Carey et al.\ 1998). Maret et al.\ (2004) observed a jump
in the abundance of gas-phase \form\ where grain mantles evaporate
($T_D=100$ K) in Class 0 protostellar envelopes, supporting the idea
of depletion onto grains.  In addition, at high density and very low
temperature, D$_2$CO forms efficiently (Tielens 1983) because of its
lower zero energy level compared to that of H$_2$CO.  D$_2$CO is
considered to form both through gas chemistry and surface chemistry
(Ceccarelli et al.\ 2002).  In dark cores without protostellar
objects, abundant deuterated molecules in the gas phase are mainly
driven by gas chemistry because the dust temperature is not high
enough for deuterated molecules on grain surfaces to be
desorbed. Gas-phase deuteration may be more efficient where CO and
NH$_3$ are depleted, as in PPCs (Roberts \& Millar 2000; Bacmann et
al.\ 2003).  D$_2$CO, which is formed and accumulated on grain
surfaces during the PPC stage, evaporates when a newly formed luminous
source heats dust grains (Loinard et al.\ 2002).

In addition to depletion, another aspect of the model results warrants
further discussion. No single model was able to fit the 6 cm and 1.3
mm data simultaneously.  We have considered different explanations for
this discrepancy.  First, we have tried four different abundance
profiles: step function, power law, exponential, and chemical
model. Table \ref{modtab} lists the model parameters for each type of
abundance profile and the corresponding absolute deviations.  The
table shows that there is not one clear best-fitting model abundance
profile for all of the sources.  The power law, exponential, and
chemical model abundance profiles are all very similar (Figure
\ref{abun}) and often result in comparable line profiles.  None of the
abundance profiles solves the problem of fitting the data at both
wavelengths. For those models providing a good match to the 6 cm
observations, the models consistently underpredict the 1.3 mm line
strength in L1498 and L1512 and show 1.3 mm self-absorption in L1544.

Second, we increased the collision rates for the transitions that
populate the upper level of the 1.3 mm transition by a factor of
three. This change did increase the model 1.3 mm emission by a factor
of 1.75 on average. However, the transitions to the upper level of the
1.3 mm transition are also significant in the pumping mechanism that
allows for 6 cm absorption against the CBR. The increase in collision
rates increased the efficiency of cooling of the 6 cm absorption, and
the absorption in the models deepened by more than a factor of
2. Therefore, a model with increased collision rates did not fit the 6
cm and 1.3 mm observed line profiles simultaneously.

Since changing the abundance profiles and collision rates did not
solve the discrepancy between the 6 cm and 1.3 mm line models, we
suggest that a density profile other than the Bonner-Ebert sphere may
be needed to simulate the observed 6 cm and 1.3 mm data. The two
transitions are most sensitive to only slightly different densities in
emission, but the 6 cm absorption can come from a region of
significantly lower density (Figure \ref{tex}). The strong 6 cm
absorption could mask weak 6 cm emission coming from the denser
regions. Therefore, we are, in effect, seeing two different parts of
the cloud with the two transitions, and a more complex density profile
may be able to match both lines. The Bonner-Ebert density profile has
been found to fit submillimeter dust continuum data well in PPCs
(Evans et al.\ 2001).  Deviations from spherical symmetry in the PPCs
not accounted for in the models may be contributing to the discrepancy
in the 6 cm and 1.3 mm models. However, a three-dimensional radiative
transfer model of the dust continuum emission from L1544 fits a
power-law density profile to the data and cannot rule out a
Bonner-Ebert density profile similar to that of the best fit
one-dimensional model (S. Doty et al., in preparation). The strength
of the 1.3 mm lines at both 0\arcsec\ and 60\arcsec\ suggests that
there is more dense material further out in the cloud than the
Bonner-Ebert or power law density profiles allow. One solution may be
to include small regions of enhanced density in the
models. High-density clumps would not be resolved in the dust
continuum and may not be large enough to affect the strength of the 6
cm line but could provide enough dense material in the outer parts of
the cloud to produce the observed 1.3 mm \form\ lines.  Additionally,
the temperature and density dependences are not as well decoupled in
the 1.3 mm transition as in the 6 cm transition.  Small regions of
higher temperature may also create the excess 1.3 mm emission
observed.

\section{Summary}\label{sec_sum}

We have presented maps of three PPCs in two transitions (6 cm and 1.3
mm) of formaldehyde showing strong absorption in the 6 cm transition
toward all the sources. The absorption extended well beyond the edge
of the observed dust emission. The 1.3 mm emission was observed out to
60\arcsec\ in all the PPCs. A velocity gradient, increasing away from
the center of the core, was observed in L1498.  The apparent gradient
may be caused by unrelated gas in the region of L1498.

We have modeled both \form\ transitions, including the offset positions,
with a multi-stage method including dust radiative transfer,
gas energetics, and gas radiative transfer. The attenuation of the ISRF,
reflected in the parameter $G_0$ in the energetics code, was
constrained by CO observations. However, since $G_0$ can have a
significant effect on the model results, full PDR analysis is needed
to better constrain this parameter and future PPC models.

The models described in this work indicate that \form\ is depleted in
the center of these PPCs as predicted by chemical network
models. Undepleted models cannot reproduce the strong 6 cm absorption
but predict 6 cm emission toward the center of the cold, dense PPCs.
Therefore, the 6 cm \form\ transition is not able to probe the
densest parts of PPCs. Although all the models require
depletion of \form, the shape of the abundance profile in the
best-fitting models varies from source to source. L1498 is fitted best
with a chemical model abundance profile that is depleted both toward
the center and at large radii. L1512 requires a large amount of
depletion out to a relatively large radius, so is fitted with a step
function abundance profile. L1544 is fitted best with a power law
abundance profile that depletes smoothly toward the center. The models
place strong constraints on the physical parameters of the core at
large radii and unequivocally indicate depletion of \form\ toward the
center of PPCs.

The models also show that multiple velocity components may be present
in the 6 cm \form\ observations of L1498 and L1512.  The 1.3 mm and
the 6 cm data cannot be well matched with a single model.  Models
that are reasonable fits to the 6 cm data are unable to reproduce the
strength of the 1.3 mm emission. The 1.3 mm emission may be probing
denser gas than the 6 cm absorption. A more complex density or
temperature distribution may be needed to fit both sets of data.

\section{Acknowledgments}

We are grateful to H. Hernandez, K. Allers, J. Wu, A. Bauer, and
M. Enoch for their long nights at Arecibo and the CSO helping to
obtain the data for this paper. This material is based in part on work
supported by the National Aeronautics and Space Administration under
Grant No. NGT5-50401 issued through the Office of Space Science
(KEY). This work was also supported by grants AST-9988230 and
AST-030725 from the National Science Foundation (NJE) and by a grant
from the Research Corporation (SDD). The National Astronomy and
Ionosphere Center is operated by Cornell University under a
Cooperative Agreement with the National Science Foundation.



\clearpage
\begin{deluxetable}{lccccccccc}
\tablecolumns{10}
\tablecaption{H$_2$CO 6 cm line results\label{ppctab}}
\tablewidth{0pt} 
\tablehead{
\colhead{Source}                &
\colhead{Offset}  &
\colhead{$v_{LSR}$}     &
\colhead{$\Delta$v}             &
\colhead{$\sigma_{turb}$} &
\colhead{$T_R$}     &
\colhead{rms}   &
\colhead{$\tau$}      &
\colhead{$T_{ex}$}    &
\colhead{N}       \\
\colhead{}                &
\colhead{(\am, \am)}  &
\colhead{(\kms)}     &
\colhead{(\kms)}             &
\colhead{(\kms)} &
\colhead{(K)}     &
\colhead{(K)}     &
\colhead{}      &
\colhead{(K)}    &
\colhead{($10^{12}$ \cmc)}                
}

\startdata 
 L1498 & (0,0)           & 7.85 & 0.35 & 0.14 &  $-$0.80  & 0.017 & 0.51 & 0.73 & 2.06 \\
       & (0,0.7)         & 7.86 & 0.31 & 0.12 &  $-$0.78  & 0.025 & 0.75 & 1.24 & 4.27 \\ 
       & (0,1.4)         & 8.02 & 0.72 & 0.30 &  $-$0.48  & 0.022 & --   & --   & -- \\
       & (0,$-$1.4)      & 7.94 & 0.37 & 0.15 &  $-$0.68  & 0.028 & 0.39 & 0.62 & 1.45 \\
       & (0.7,0)         & 7.88 & 0.36 & 0.14 &  $-$0.73  & 0.023 & 0.69 & 1.26 & 4.62 \\ 
       & (0.7,0.7)       & 7.97 & 0.64 & 0.27 &  $-$0.45  & 0.033 & 0.31 & 1.03 & 3.05 \\
       & (0.7,2.1)       & 8.34 & 0.95 & 0.40 &  $-$0.18  & 0.020 & --   & --   & -- \\
       & (0.7,$-$0.7)    & 7.92 & 0.56 & 0.23 &  $-$0.75  & 0.030 & --   & --   & -- \\
       & (1.4,0)         & 7.98 & 0.60 & 0.25 &  $-$0.55  & 0.023 & 0.37 & 0.95 & 3.20 \\
       & (1.4,1.4)       & 8.17 & 0.92 & 0.39 &  $-$0.30  & 0.032 & --   & --   & -- \\
       & (2.1,0.7)       & 8.16 & 0.80 & 0.34 &  $-$0.28  & 0.023 & --   & --   & -- \\
       & ($-$0.7,0.7)    & 7.85 & 0.28 & 0.11 &  $-$0.77  & 0.030 & 1.02 & 1.53 & 6.34 \\
       & ($-$0.7,$-$0.7) & 7.90 & 0.30 & 0.12 &  $-$0.68  & 0.025 & 1.31 & 1.66 & 7.43 \\
       & ($-$0.7,$-$2.1) & 8.03 & 0.37 & 0.15 &  $-$0.43  & 0.025 & --   & --   & --  \\ 
       & ($-$1.4,0)      & 7.89 & 0.28 & 0.11 &  $-$0.90  & 0.020 & 0.87 & 1.17 & 4.24 \\
       & ($-$1.4,$-$1.4) & 8.01 & 0.54 & 0.22 &  $-$0.43  & 0.028 & 0.53 & 1.68 & 6.97 \\
       & ($-$2.1,$-$0.7) & 7.92 & 0.34 & 0.13 &  $-$0.60  & 0.023 & 0.48 & 1.15 & 2.77 \\
 L1512 & (0,0)           & 7.11 & 0.26 & 0.10 &  $-$0.73  & 0.012 & 1.31 & 1.72 & 8.44 \\
       & (0,1)           & 7.06 & 0.26 & 0.10 &  $-$0.65  & 0.025 & 1.30 & 1.83 & 8.89 \\ 
       & (0,$-$1)        & 7.13 & 0.23 & 0.08 &  $-$0.77  & 0.032 & 1.23 & 1.64 & 6.71 \\ 
       & (1,0)           & 7.11 & 0.26 & 0.10 &  $-$0.75  & 0.027 & 1.15 & 1.63 & 7.03 \\ 
       & (1,1)           & 7.09 & 0.26 & 0.10 &  $-$0.80  & 0.025 & 0.99 & 1.45 & 5.49 \\
       & (1,$-$1)        & 7.13 & 0.23 & 0.08 &  $-$0.65  & 0.023 & 1.45 & 1.88 & 8.99 \\
       & (2,0)           & 7.15 & 0.30 & 0.12 &  $-$0.52  & 0.023 & 1.00 & 1.91 & 8.19 \\
       & (2,1)           & 7.16 & 0.54 & 0.22 &  $-$0.38  & 0.022 & --   & --   &  -- \\ 
       & (2,$-$1)        & 7.22 & 0.49 & 0.20 &  $-$0.22  & 0.018 & --   & --   &  -- \\ 
       & ($-$1,0)        & 7.12 & 0.27 & 0.10 &  $-$0.80  & 0.025 & 1.17 & 1.56 & 7.17 \\ 
       & ($-$1,1)        & 7.06 & 0.25 & 0.09 &  $-$0.75  & 0.018 & 1.48 & 1.75 & 9.37 \\
       & ($-$1,$-$1)     & 7.12 & 0.25 & 0.09 &  $-$0.80  & 0.027 & 0.55 & 0.84 & 1.80 \\
       & ($-$2,0)        & 7.12 & 0.28 & 0.11 &  $-$0.62  & 0.025 & 1.00 & 1.75 & 7.05 \\
       & ($-$2,1)        & 7.06 & 0.25 & 0.09 &  $-$0.70  & 0.027 & 1.03 & 1.63 & 6.06 \\ 
       & ($-$2,$-$1)     & 7.11 & 0.27 & 0.10 &  $-$0.55  & 0.027 & 0.64 & 1.56 & 3.92 \\ 
 L1544 & (0,0)           & 7.21 & 0.31 & 0.12 &  $-$0.87  & 0.012 & 2.17 & 1.75 & 16.9 \\
       & (0,1.4)         & 7.14 & 0.38 & 0.15 &  $-$0.92  & 0.015 & 1.85 & 1.64 & 16.6 \\
       & (0,$-$1.4)      & 7.27 & 0.28 & 0.11 &  $-$0.88  & 0.022 & 1.69 & 1.64 & 11.2 \\
       & (0.7,0.7)       & 7.14 & 0.43 & 0.17 &  $-$0.82  & 0.033 & 2.01 & 1.78 & 22.2 \\
       & (0.7,2.1)       & 7.03 & 0.35 & 0.14 &  $-$1.05  & 0.022 & 2.13 & 1.53 & 16.6 \\
       & (0.7,$-$0.7)    & 7.21 & 0.29 & 0.11 &  $-$1.05  & 0.022 & 2.00 & 1.51 & 12.8 \\
       & (1.4,0)         & 7.15 & 0.32 & 0.13 &  $-$1.12  & 0.017 & 1.53 & 1.30 & 9.35 \\
       & (1.4,1.4)       & 7.08 & 0.40 & 0.16 &  $-$0.87  & 0.027 & 2.16 & 1.74 & 21.7 \\
       & (2.1,0.7)       & 7.13 & 0.29 & 0.11 &  $-$1.25  & 0.022 & 1.68 & 1.19 & 8.58 \\
       & ($-$0.7,0.7)    & 7.23 & 0.33 & 0.13 &  $-$0.98  & 0.023 & 2.08 & 1.60 & 15.9 \\
       & ($-$0.7,$-$0.7) & 7.25 & 0.27 & 0.10 &  $-$1.07  & 0.023 & 1.60 & 1.39 & 8.79 \\
       & ($-$0.7,$-$2.1) & 7.27 & 0.30 & 0.12 &  $-$0.67  & 0.027 & 1.37 & 1.83 & 10.8 \\
       & ($-$1.4,0)      & 7.24 & 0.30 & 0.12 &  $-$1.10  & 0.027 & 2.03 & 1.46 & 13.0 \\
       & ($-$1.4,$-$1.4) & 7.26 & 0.32 & 0.13 &  $-$0.68  & 0.027 & 1.37 & 1.81 & 11.3 \\
       & ($-$2.1,$-$0.7) & 7.28 & 0.25 & 0.09 &  $-$0.63  & 0.017 & 1.49 & 1.91 & 10.1 \\

\enddata

\end{deluxetable}

\clearpage
\begin{deluxetable}{lccccccc}
\tablecolumns{8}
\tablecaption{\form\ 1.3 mm Line Results \label{csotab}}
\tablewidth{0pt} 
\tablehead{
\colhead{Source}                &
\colhead{Offset}       &
\colhead{$v_{LSR}$}         &
\colhead{$\Delta$v}             &
\colhead{$\sigma_{turb}$}               &
\colhead{$T_R$}             &
\colhead{rms}                  &
\colhead{$\eta_{MB}$}    \\
\colhead{}                      &
\colhead{(\am, \am)}  &
\colhead{(km s$^{-1}$)}     &    
\colhead{(km s$^{-1}$)}          &
\colhead{(km s$^{-1}$)}      &
\colhead{(K)}    &  
\colhead{(K)}    &  
\colhead{}
 }

\startdata 

 L1498   & (0,0)           & 7.82 & 0.23 & 0.08 & 0.23  & 0.043 & 0.53 \\
	 & (0.7,0.7)       & 7.85 & 0.12 &  --  & 0.21  & 0.056 & 0.77 \\
	 & (0.7,$-$0.7)    & 7.82 & 0.35 & 0.14 & 0.14  & 0.043 & 0.77 \\ 
	 & ($-$0.7,0.7)    &  --  &  --  & --   & $<$ 0.16\tablenotemark{a} & 0.081 & 0.77 \\
	 & ($-$0.7,$-$0.7) & 7.85 & 0.25 & 0.09 & 0.23  & 0.056 & 0.77 \\   
 L1512   & (0,0)           & 7.01 & 0.35 & 0.14 & 0.34  & 0.079 & 0.53 \\
	 & (0,1)           &  --  &  --  &  --  & $<$ 0.10\tablenotemark{a} & 0.051 & 0.77 \\
	 & (0,$-$1)        & 7.17 & 0.27 & 0.10 & 0.34  & 0.070 & 0.77 \\
	 & (1,0)           & 7.09 & 0.31 & 0.12 & 0.17  & 0.029 & 0.77 \\
	 & ($-$1,0)        & 7.12 & 0.13 & 0.02 & 0.18  & 0.034 & 0.77 \\ 
 L1544   & (0,0)           & 7.14 & 0.41 & 0.17 & 0.47  & 0.063 & 0.57 \\
	 & (0.7,0.7)       & 7.06 & 0.48 & 0.20 & 0.23  & 0.051 & 0.77 \\
	 & (0.7,$-$0.7)    & 7.27 & 0.43 & 0.17 & 0.22  & 0.056 & 0.77 \\
         & ($-$0.7,0.7)    & 7.12 & 0.44 & 0.18 & 0.45  & 0.088 & 0.77 \\
         & ($-$0.7,$-$0.7) & --   &  --  & --   & $<$ 0.12\tablenotemark{a} & 0.058 & 0.77 \\

\enddata

\tablenotetext{a}{2-$\sigma$ upper limit}

\end{deluxetable}

 
\begin{deluxetable}{ccccccccc}
\tablecolumns{9}
\tablecaption{Model Parameters \label{modtab}}
\tablewidth{0pt} 
\tablehead{
\colhead{Source}                &
\colhead{Abun.}     &
\colhead{$X_0$}       &
\colhead{$r_D$}             &
\colhead{Other}         &
\multicolumn{2}{c}{Abs. Dev. (6 cm)}         &
\multicolumn{2}{c}{Abs. Dev. (1.3 mm)}           \\
\colhead{}                      &
\colhead{Prof.}            &
\colhead{} &
\colhead{(pc)}     &    
\colhead{}          &
\colhead{Int.\tablenotemark{a}}          &
\colhead{Prof.\tablenotemark{b}}  &
\colhead{Int.\tablenotemark{a}}          &          
\colhead{Prof.\tablenotemark{b}}        
}

\startdata 
 L1498 & step  & 4\ee{-9} & 0.02 & $f_D$=10 & 0.17 & 1.28 & 0.053 & 0.29    \\
       & power & 7\ee{-9}  & --   & $p$=1\tablenotemark{c}    & 0.25 & 1.43 & 0.055 & 0.30    \\
       & exponential  & 5\ee{-9} & 0.02 & -- & 0.17 & 1.28 & 0.055 & 0.30     \\
       & chemical\tablenotemark{d} & 0.8\tablenotemark{e} & -- & $A_V$=2, $n_c$=$10^7$\cmv \tablenotemark{f} & 0.19 & 1.24 & 0.037 & 0.25      \\
       & chemical & 2.0\tablenotemark{e} & -- & $A_V$=2, $n_c$=$10^4$\cmv \tablenotemark{f}, $G_0$=0.6 & 0.26 & 1.27 & 0.058 & 0.30      \\

 L1512 & step\tablenotemark{d}  & 1\ee{-8} & 0.05 & $f_D$=20 & 0.037 & 0.96 & 0.10  & 0.71     \\
       & power & 1.5\ee{-8} & -- & $p$=1\tablenotemark{c}    & 0.081 & 1.04 & 0.078 & 0.65    \\
       & exponential & 1.5\ee{-8} & 0.05 & -- & 0.079 & 1.24 & 0.050 & 0.60    \\
       & chemical & 0.3\tablenotemark{e} & -- & $A_V$=3, $n_c$=$10^6$\cmv \tablenotemark{f} & 0.13 & 0.92 & 0.078 & 0.67    \\

 L1544 & step   & 1.5\ee{-8} & 0.04 & $f_D$=10 & 0.13 & 1.07 & 0.15 & 0.97  \\
       & power  & 2.5\ee{-8} & --   & $p$=1\tablenotemark{c}    & 0.09 & 0.91 & 0.21 & 1.12  \\
       & pow. + vel.\tablenotemark{d} & 2.5\ee{-8} & -- & $p$=1, Plummer vel. & 0.084 & 0.91 & 0.23 & 1.17 \\
       & exponential   & 2\ee{-8} & 0.03 & --  & 0.11 & 0.99 & 0.20 & 1.13  \\
       & chemical  & 0.5\tablenotemark{e} & -- & $A_V$=3, $n_c$=$10^6$\cmv \tablenotemark{f} & 0.11 & 0.89 & 0.20 & 1.10  \\
      
\tablenotetext{a}{Average absolute deviation of the integrated intensity of the model versus the observed lines over all offsets.} 
\tablenotetext{b}{Average absolute deviation of the modeled profile shape versus the observed profiles over all offsets.}
\tablenotetext{c}{Power law exponent.}
\tablenotetext{d}{Best-fit model described in text and shown in Figures \ref{l1498mod} -- \ref{l1544mod}.}
\tablenotetext{e}{Since the chemical models are not characterized by
$X_0$, this value indicates the scale factor for the chemical model
abundance profile.}  
\tablenotetext{f}{These values for $A_V$ and $n_c$ describe the input
for the chemical model and are independent of dust radiative
transfer model values.}

\enddata

\end{deluxetable}


\clearpage

\appendix

\section{APPENDIX: Energetics}

\begin{figure}
\plotone{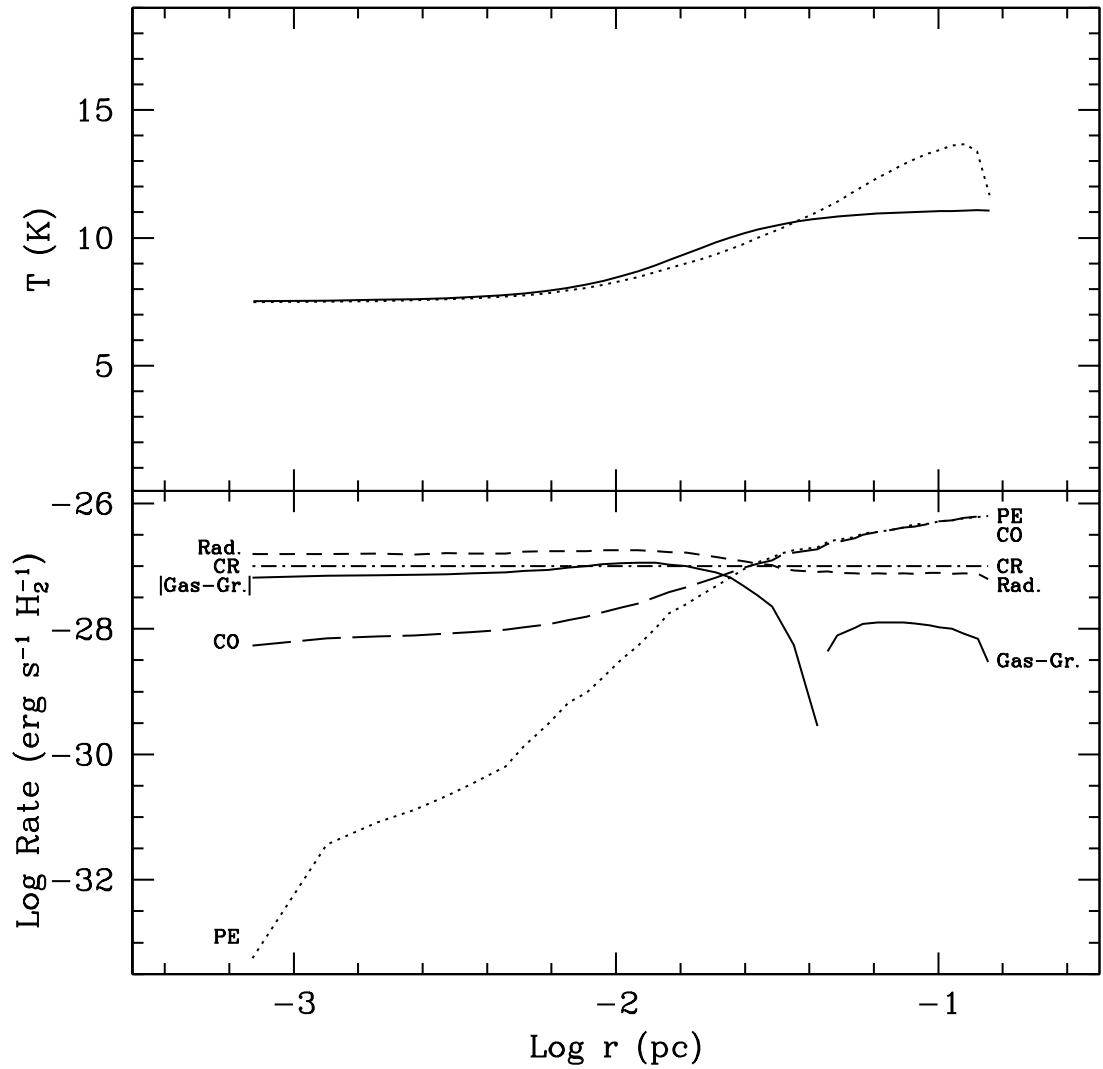}\figcaption{\label{rates} The top panel shows the
dust temperature (solid line) from the dust radiative transfer model
and the gas temperature (dotted line) produced by the energetics code
for a model of L1544. The bottom panel shows the heating and cooling
rates per hydrogen molecule in the energetics code. The lines
represent, from top to bottom on the left side, the total gas
radiative cooling, cosmic ray heating, the absolute value of the
gas-grain energy exchange, CO radiative cooling, and the photoelectric
effect.  The break in the solid line represents the point at which the
gas-grain energy transfer is no longer heating the gas. At high
density and low temperature, in the inner parts of the PPC, the total
gas radiative cooling is large compared to CO cooling due mainly to
the contributions of isotopologues of CO.}
\end{figure}

The gas temperature at each depth in the model is determined by
balancing the local heating and cooling rates (Figure \ref{rates}).  The
approach used is patterned after that of Doty \& Neufeld (1997), but
with updates that we describe here.

Before discussing the detailed heating and cooling rates, we note that
all references to $n(\mathrm{H}_{2})$ correspond to H$_{2}$ number
density.  However, many of the adopted rates require the baryon number
density, $n(\mathrm{H}) \sim 2n(\mathrm{H}_{2}) + 4n(\mathrm{He})$
(assuming that there is only neutral and molecular gas, and low
abundances of heavier elements).  For a molecular gas with 25\% helium
by mass, $n({\mathrm{He}}) = 0.166 n(\mathrm{H}_{2})$, and
$n({\mathrm{H}}) = 2.66 n(\mathrm{H}_{2})$.  All expressions below
have accounted for this fact.

The gas close to the surface is primarily heated by photoelectric
heating.  We follow the prescription of Bakes \& Tielens (1994),
namely
\begin{equation}
\Gamma_{\mathrm{pe}} = 10^{-24} \epsilon G(r) n(\mathrm{H})
\;\mathrm{ ergs} \;\mathrm{ cm}^{-3} \;\mathrm{ s}^{-1},
\end{equation}
where the efficiency, $\epsilon$ is given by
\begin{equation}
\epsilon = \frac{4.87\times10^{-2}}{1+4\times
10^{-3}(G(r)T_K^{0.5}/n_{e})^{0.73}}
+
\frac{3.65\times10^{-2}(T_K/10^{4})^{0.7}}{1+2\times10^{-4}(G(r)T_K^{0.5}/n_{e})}.
\end{equation}
In this expression, $T_K$ is the temperature, $n_{e}$ is the electron
number density, and $G(r) = \overline{J}(r;E>6\mathrm{eV}) /
\overline{J}(\mathrm{ISRF};E>6\mathrm{eV})$ is the ratio of the mean
intensity of the local radiation field within the cloud to the mean
intensity of the ISRF for energies above 6 eV -- the energies above
which photons can cause significant photoelectric heating (e.g.,
Juvela, Padoan, \& Jimenez 2003).  The electron number density is
taken from chemical models similar to Doty et al. (2002), which
include UV processing and attenuation.  In the clouds we study, the
product $G(r)T_K^{0.5}/n_{e}$ is small enough that the total heating
rate scales with $G(r)$ and is nearly independent of the exact value
of the electron number density.

The depth-dependent radiation field is determined by a self-consistent
solution to the radiative transfer problem using the code of Egan,
Leung, \& Spagna (1988).  In this case, we assume spherical symmetry,
include scattering, and adopt the dust properties for coagulated
grains with icy mantles described by Ossenkopf \& Henning (1994;
column (5) of Table 1).  These assumptions are identical to those
adopted for the dust radiative transfer models, providing a consistent
approach.  While the radiation field is a complicated function of
depth, it is interesting to note that it is generally much stronger
than would be calculated by taking a simple attenuation law used for
plane-parallel slabs such as $G(r) = G_0\mathrm{exp}(-a\tau_{V})$,
where $\tau_{V}$ is the extinction at V and $a$ is commonly taken to
be in the range of 1.8 (e.g., Tielens, private communication) to 2.5
(Hollenbach, Werner, \& Salpeter 1971).  In a spherical geometry, any
point in space has a lower optical depth to the outside for non-normal
rays than for the equivalent ray in a slab geometry.  As a result, the
radiation field along non-normal rays can penetrate further into
spherical sources, leading to a higher radiation field (e.g., Flannery
et al. 1980).  In the models described in the text, the value of $G_0$
is varied until agreement with the CO $J=1-0$ line is reproduced.
Physically, this has the effect of accounting for variations in the
incident radiation field due to local sources and/or attenuation by
the surrounding low-density cloud.

Grain photoelectric heating usually dominates in the exterior of the
cloud.  However, due to attenuation of the photon flux by the dust,
cosmic ray heating will often dominate in the interior ($\tau_{V} >
1$), as can be seen in Figure \ref{rates}.  We assume a cosmic-ray
ionization rate of $3 \times 10^{-17}$ s$^{-1}$ (van der Tak \& van
Dishoeck 2000), and also take the energy input per ionization to be
$\Delta Q = 20$ eV (Goldsmith 2001).  This yields a cosmic ray heating
rate per unit volume of
\begin{equation}
\Gamma_{\mathrm{gas,cr}} = 10^{-27} n(\mathrm{H}_{2})
\left [\frac{\zeta}{3\times10^{-17}\mathrm{s}^{-1}}\right ]
\left [\frac{\Delta Q}{20\mathrm{eV}}\right ]
\;\mathrm{ ergs} \;\mathrm{ cm}^{-3} \;\mathrm{ s}^{-1}.
\end{equation}

For the gas cooling, we follow Doty \& Neufeld (1997) in their
adoption of the tabulated cooling functions of Neufeld, Lepp, \&
Melnick (1995) and Neufeld \& Kaufman (1993) for CO, OI, H$_{2}$,
H$_{2}$O, and other diatomic and polyatomic molecules.  As discussed
in the text, molecular depletion is a key aspect of these
clouds. However we have not decreased the abundances of coolants
because the abundances are strongly affected only at relatively high
densities, where $T_K$ is already closely coupled to $T_d$.  Depletion
has only a small effect on gas temperature (Goldsmith 2001).  Similar
to Doty \& Neufeld (1997), we find that the cooling in the exterior is
dominated by CO.  Likewise, the decrease in temperature on the outside
of the models is due to the increased ability of radiation to escape,
allowing the gas to cool.  We tested the case where the gas
temperature is constant at the outer radii rather than decreasing and
found that the drop off in $T_K$ at large radii causes a $\sim$5\%
decrease in $T_R$.

Collisions between gas and dust can either heat or cool the gas; the
dust will cool the gas if the gas kinetic temperature, $T_K$, is
greater than the dust temperature, $T_d$, and vice-versa. Our
formulation treats the interaction as a cooling term formally.  There
is of course a balancing heating or cooling of the dust in the
process, but the effect on $T_d$ is small compared to radiative
processes, even deep in well-shielded clouds (Evans et al.  2001,
Goldsmith 2001), and we ignore it here. We adopt the general
prescription of Hollenbach \& McKee (1989), assuming collisions
between gas and dust where the dust has a power-law size distribution
$n(a) \propto a^{-3.5}$ (Mathis, Rumpl, \& Nordsieck 1977), with a
minimum grain size $a_{\mathrm{min}}=100$ Angstroms, a maximum grain
size $a_{\mathrm{max}}=0.25$ $\mu$m, a mass density $\rho = 2$ g
cm$^{-3}$, and a dust-to-gas mass ratio of $R_{\mathrm{dg}}=4.86
\times 10^{-3}$.  These assumptions lead to an average dust cross
section per baryon of $\Sigma_{\mathrm{d}}=6.09\times10^{-22}$
cm$^{-2}$.  We adopt the temperature-dependent accommodation
coefficient $\alpha = 0.37 [1 -
0.8\mathrm{exp}(-75/T_{\mathrm{K}})]$ (Burke \& Hollenbach 1983). This
expression comes from averaging over speeds for an astrophysical
mixture of gases; thus it must be used in conjunction with a mean
speed for a hydrogen atom, \mean{v(m_H)}.  With these conventions, our
adopted expression for the gas-dust energy transfer rate is
\begin{equation}
\Lambda_{\mathrm{gd}}=9.0\times10^{-34} n(\mathrm{H})^{2}
(T_{\mathrm{K}})^{0.5}
\left (1 - 0.8\times\mathrm{exp}\left (-\frac{75}{T_{\mathrm{K}}}\right )\right )
(T_{\mathrm{K}}-T_{\mathrm{d}})
\left [\frac{\Sigma_{\mathrm{d}}}{6.09\times10^{-22}}\right ]
\;\mathrm{ ergs} \;\mathrm{ cm}^{-3} \;\mathrm{ s}^{-1}.
\end{equation}

This expression is similar in form to others in the literature.
However, the coefficient is approximately the geometrical mean of the
lower value of Goldsmith (2001) and the higher value of Hollenbach \&
McKee (1989).  It is, however, within a factor of 5 of these two other
values.  This is reassuring as our expression does not represent an
extremum on either end of the range of expected gas-dust cooling
rates, and agrees to within the intrinsic uncertainties in assumed
microphysical properties (e.g., dust composition, size distribution,
dust-gas mass ratio, sticking probabilities, gas composition, etc.).
In practice, gas-dust cooling has the effect, as noted by other
authors (e.g., Doty \& Neufeld 1997; Goldsmith 2001), of strongly
coupling the gas and dust temperatures at densities $n > 10^{5}$
cm$^{-3}$.

These heating and cooling rates are then solved for the temperature
point-by-point within the cloud.  The problem is treated as a root-finding
problem in the gas temperature, and solved using Brent's method
(Press et al. 1992) to a fractional accuracy of $10^{-8}$.


\end{document}